\documentclass[11pt]{article}
\usepackage[utf8]{inputenc}
\usepackage[T1]{fontenc}
\usepackage{lmodern}
\usepackage{amsmath,amssymb,amsthm,mathtools}
\usepackage[margin=1in]{geometry}
\usepackage[round,authoryear]{natbib}
\usepackage{booktabs,array,enumitem,microtype,xcolor}
\usepackage{caption}
\usepackage{algorithm}
\usepackage{algpseudocode}
\usepackage[colorlinks=true,linkcolor=blue!50!black,citecolor=blue!50!black,urlcolor=blue!50!black]{hyperref}

\newtheorem{theorem}{Theorem}
\newtheorem{proposition}{Proposition}
\newtheorem{lemma}{Lemma}
\newtheorem{corollary}{Corollary}
\theoremstyle{definition}

\newtheorem{example}{Example}

\newcommand{\R}{\mathbb{R}}
\newcommand{\N}{\mathrm{N}}
\newcommand{\Ga}{\mathrm{Gamma}}

\newcommand{\rank}{\operatorname{rank}}
\newcommand{\tr}{\operatorname{tr}}
\newcommand{\diag}{\operatorname{diag}}
\newcommand{\Var}{\operatorname{Var}}
\newcommand{\E}{\mathrm{E}}
\newcommand{\norm}[1]{\lVert #1\rVert}
\newcommand{\Deta}{D_\eta}
\newcommand{\pe}{\preceq}
\newcommand{\ind}{\stackrel{\text{ind}}{\sim}}

\title{Geometric ergodicity of Gibbs samplers for linear mixed models with several heavy-tailed random-effect components and unknown tail parameters}
\author{Rolando de la Cruz\thanks{Faculty of Engineering and Sciences, Universidad Adolfo Ib\'a\~nez, Pe\~nalol\'en, Chile; Data Observatory Foundation, Santiago, Chile; Millennium Nucleus for Social Data Science (SODAS), Santiago, Chile. Corresponding author: \texttt{rolando.delacruz@uai.cl}}
\and Nathalie Humeniy\thanks{Departamento de Estad\'istica, Pontificia Universidad Cat\'olica de Chile, Santiago, Chile.}}
\date{\today}

\begin{document}
\maketitle

\begin{abstract}
Linear mixed models with heavy-tailed random effects are widely used for robust Bayesian analysis of longitudinal data. We study the three-block Gibbs sampler for such models when the random effects are split into several components, such as random intercepts and slopes, each with its own scale and its own scale mixture of normals, when the tail parameters have priors on finite grids and are drawn jointly with the latent weights, and when the random-effects design is rank deficient, as when some subjects contribute a single observation. We give sufficient conditions for geometric ergodicity that attribute the rank deficiency to the components that cause it, and show that, under conditions on the gamma priors on the precision parameters, they hold for Student-t and slash random effects with individual weights, any grid of tail parameters and any design. The result extends to random-effect vectors whose coordinates share a weight, for subject-specific designs with known shape matrices. It justifies Monte Carlo standard errors for regression coefficients, log-scales and posterior probabilities of the tail parameters. Simulations show that the scale and the tail parameter can mix much more slowly than the regression coefficients, and an analysis of beta-HCG data illustrates the methods.
\end{abstract}

\noindent\textbf{Keywords:} drift condition; heavy-tailed random effects; multiple variance components; rank-deficient design; Student-$t$ degrees of freedom; Markov chain central limit theorem.

\section{Introduction}\label{sec:intro}

Linear mixed models with heavy-tailed random effects are widely used to accommodate subjects whose trajectories depart from the bulk of the population \citep{Lange1989,Pinheiro2001,Rosa2003,Lachos2010}. In a Bayesian analysis the random effects are usually represented as scale mixtures of normals: each random effect is normal given a latent precision weight, and the weights follow a mixing distribution $H$, such as the gamma law that produces Student-$t$ random effects. With conditionally conjugate priors this representation gives a Gibbs sampler with three blocks: the variance components, the latent weights, and the regression coefficients together with the random effects.

Longitudinal applications often combine three features that make the convergence of this sampler harder to analyse: several scales, uncertainty about the tails, and few observations per subject. Their interaction is the subject of this paper.

First, the random effects are split into $K$ components, for instance random intercepts and random slopes, or the effects of different groups of subjects, each with its own scale parameter and its own mixing distribution, and the random-effects design matrix $Z$ is rank deficient. For a random intercept and slope, every subject with a single observation produces two proportional columns of $Z$; in the $\beta$-HCG data of Section~\ref{sec:app}, 52 of 173 women have a single measurement and $\rank(Z)=q-52$. With several scales and heavy tails, the latent weights interact with the null space of $Z$ component by component, and the rank deficiency must be attributed to the components that cause it rather than charged to all of them.

Second, the tail parameter of each mixing distribution is rarely known. Predictive criteria can be used to choose it \citep{delacruz:humeniy:2026}, but a fully Bayesian treatment places a prior on it \citep{Fonseca2008,VillaWalker2014}. A continuous prior may require an additional nonconjugate update. With a prior on a finite grid, the tail parameter can be drawn exactly and jointly with the weights, and the algorithm remains a three-block Gibbs sampler. We study this version.

Third, in robust random-coefficient models the coordinates of a subject's random-effect vector usually share one weight \citep{Pinheiro2001,Lachos2010}, which changes the dependence between the weights and the random effects.

Geometric ergodicity, together with a moment condition, guarantees a Markov chain central limit theorem (CLT) and hence valid Monte Carlo standard errors \citep{Jones2004,fleg:hara:jone:2008,vats:fleg:jone:2019}. For normal random effects, \citet{roma:hobe:2012,roma:hobe:2014} gave conditions for geometric ergodicity of this sampler with several variance components and arbitrary $Z$, and \citet{abra:hobe:2017} later removed some of their assumptions. For random effects with a known mixing distribution, \citet{delacruz:humeniy:2026} established geometric ergodicity for a single scale parameter and every design, and for scales indexed by a known grouping when $Z$ has full column rank. Rank-deficient designs with several scales, unknown tail parameters and shared weights are not covered by these results.

Our main results are as follows. Theorem~\ref{thm:main} gives sufficient conditions for geometric ergodicity with $K$ components, arbitrary $Z$ and tail parameters on finite grids. The conditions on the shapes of the gamma priors on the precision parameters involve, for each component $k$, the quantity $\delta_k=q_k+r_{-k}-r$, where $r=\rank(Z)$ and $r_{-k}$ is the rank of $Z$ without its $k$th block, so that only the components that contribute to the rank deficiency are restricted; for normal random effects these conditions have the same form as those of \citet{roma:hobe:2014}, and are no stronger, and potentially strictly weaker, than theirs when $K>1$. The key ingredient is a bound for the block sums of the diagonal of a projection onto a rescaled null space (Lemma~\ref{lem:blocklev}), which attributes the rank deficiency component by component even though the weights rescale the null space at every step. The condition on the mixing distributions is expressed through the projection onto the null space of $Z$, and Proposition~\ref{prop:families} translates it for Student-$t$ and slash random effects with unknown tail parameters. Adapting the rescaling argument of \citet{delacruz:humeniy:2026} to several components and unknown tail parameters (Corollaries~\ref{cor:scaling} and~\ref{cor:anydesign}), the restriction on the tails disappears for all the usual families (Corollary~\ref{cor:families}). Theorem~\ref{thm:block} allows the coordinates of a subject's random-effect vector to share one weight, under a condition that combines the tilted laws of the shared weights with the geometry of the blocks of the null space, and Corollary~\ref{cor:mvt} deduces geometric ergodicity for multivariate Student-$t$ and slash random effects with known shape matrices, unknown group-specific scales, any finite grids of tail parameters and any subject-specific design.

For Bayesian practice, the results justify Monte Carlo standard errors for the regression coefficients, the logarithms of the scale parameters and the posterior probabilities of the tail parameters, which quantify the evidence for heavy tails component by component. Section~\ref{sec:sim} shows why these errors should be computed for each quantity separately: with shared heavy-tailed weights, the scale of the random effects and the tail parameter can mix several times more slowly than the regression coefficients, and the multivariate criterion, based on the multivariate effective sample size, does not ensure a prescribed Monte Carlo precision for each component.

The rest of the paper is organized as follows. Section~\ref{sec:model} describes the model and the sampler. Section~\ref{sec:results} states the results, and Section~\ref{sec:proofs} contains the proofs. Section~\ref{sec:app} applies the results to the $\beta$-HCG data, Section~\ref{sec:sim} studies the exploration of the scale and the tail parameter with shared weights on simulated data, and Section~\ref{sec:disc} discusses extensions. Appendix~\ref{app:code} describes how the implementation was verified, and Appendix~\ref{app:lemmas} proves the auxiliary lemmas.

\section{Model and Gibbs sampler}\label{sec:model}

\subsection{Model}
Let $y$ be an $N\times1$ response, $X$ an $N\times p$ fixed-effects design and $Z=(Z_1\ \cdots\ Z_K)$ a random-effects design whose block $Z_k$ is $N\times q_k$; let $q=\sum_kq_k$ and, for $i=1,\dots,q$, let $k(i)$ be the component to which coordinate $i$ belongs. Each component $k$ has a scale parameter $\lambda_k$ and a tail parameter $d_k$ taking values in a finite set $\mathcal D_k\subset\R$. For each coordinate $i$ and each $d\in\mathcal D_{k(i)}$, let $H_{i,d}$ be a probability distribution on $\R_+=(0,\infty)$. We consider
\begin{align}
Y\mid\beta,u,\lambda_e&\sim\N_N(X\beta+Zu,\lambda_e^{-1}I_N),\label{eq:model-a}\\
u_i\mid\lambda,\eta,d&\ind\N\big(0,(\lambda_{k(i)}\eta_i)^{-1}\big),\qquad \eta_i\mid d\ind H_{i,d_{k(i)}},\qquad i=1,\dots,q,\label{eq:model-b}\\
\beta&\sim\N_p(\mu_\beta,\Sigma_\beta),\quad\lambda_e\sim\Ga(a_e,b_e),\quad\lambda_k\sim\Ga(a_k,b_k),\quad d_k\sim\pi_k,\label{eq:model-c}
\end{align}
with $\beta$, $\lambda_e$, $\lambda_1,\dots,\lambda_K$ and $d_1,\dots,d_K$ independent a priori, $\Sigma_\beta$ positive definite, all gamma shapes and rates positive, and $\pi_k$ a probability on $\mathcal D_k$ with $\pi_k(d)>0$ for every $d\in\mathcal D_k$. Here $\Ga(a,b)$ has density proportional to $x^{a-1}e^{-bx}$. A known tail parameter corresponds to a singleton $\mathcal D_k$. In most applications $H_{i,d}=H_{k(i),d}$ does not depend on the coordinate within a component; we allow coordinate-specific laws because the rescaling argument of Section~\ref{sec:scaling} produces them.

Given $\lambda_{k(i)}$ and $d_{k(i)}$, $u_i$ is a scale mixture of normals. Common families, parametrized through the precision multiplier $\eta$, are the Student-$t$ law with $d$ degrees of freedom, $H_d=\Ga(d/2,d/2)$; the slash law with parameter $\nu$, $H_\nu=\mathrm{Beta}(\nu,1)$; the contaminated normal, normal-generalized inverse Gaussian and variance-gamma laws \citep[Table~1]{delacruz:humeniy:2026}. The posterior is proper, because the likelihood is bounded by $(\lambda_e/2\pi)^{N/2}$, which is integrable against the gamma prior of $\lambda_e$, and all the other priors are proper.

We write $\lambda=(\lambda_e,\lambda_1,\dots,\lambda_K)$, $d=(d_1,\dots,d_K)$, $\theta=(\beta^T,u^T)^T$, $W=(X\ Z)$ and
\[
\Deta=\diag(\eta_1,\dots,\eta_q),\qquad\Lambda_{\lambda,\eta}=\diag(\lambda_{k(1)}\eta_1,\dots,\lambda_{k(q)}\eta_q).
\]

\subsection{Full conditionals}
For a probability $H$ on $\R_+$ with $m_{1/2}(H)=\int\eta^{1/2}H(d\eta)<\infty$ and $s\ge0$, let
\begin{equation}\label{eq:tilted}
\mathcal Z_H(s)=\int_0^\infty\eta^{1/2}e^{-s\eta}H(d\eta),\qquad H_s(d\eta)=\frac{\eta^{1/2}e^{-s\eta}H(d\eta)}{\mathcal Z_H(s)} .
\end{equation}
Collecting the factors of the joint posterior density that involve each block gives the following full conditionals.
\begin{enumerate}[label=(\roman*),leftmargin=*]
\item Given $(d,\eta,\theta)$, the components of $\lambda$ are independent, with $\lambda_e\sim\Ga\big(a_e+N/2,\,b_e+\norm{y-W\theta}^2/2\big)$ and $\lambda_k\sim\Ga\big(a_k+q_k/2,\,b_k+\sum_{i:k(i)=k}\eta_iu_i^2/2\big)$. This law does not depend on $d$.
\item Given $(\lambda,\theta)$, the pairs $(d_k,\{\eta_i\}_{i:k(i)=k})$, $k=1,\dots,K$, are independent. With $s_i=\lambda_{k(i)}u_i^2/2$,
\begin{equation}\label{eq:dpost}
\pi(d_k\mid\lambda,\theta)\propto\pi_k(d_k)\prod_{i:k(i)=k}\mathcal Z_{H_{i,d_k}}(s_i),
\end{equation}
and, given $d_k$, the $\eta_i$ with $k(i)=k$ are independent with laws $(H_{i,d_k})_{s_i}$. Since $\mathcal D_k$ is finite, $d_k$ is drawn exactly from \eqref{eq:dpost} and then the weights from their tilted laws. For the Student-$t$ family, $\mathcal Z_{H_d}(s)=(d/2)^{d/2}\Gamma(\frac{d+1}2)\Gamma(\frac d2)^{-1}(d/2+s)^{-(d+1)/2}$ and $(H_d)_s=\Ga\big(\frac{d+1}2,\frac d2+s\big)$; for the slash family, $\mathcal Z_{H_\nu}(s)=\nu\,\Gamma(\nu+\frac12)P(\nu+\frac12,s)\,s^{-(\nu+1/2)}$ for $s>0$, with $P$ the regularized lower incomplete gamma function, and $(H_\nu)_s$ is the $\Ga(\nu+\frac12,s)$ law truncated to $(0,1)$; for $s=0$, $\mathcal Z_{H_\nu}(0)=\nu/(\nu+\frac12)$ and $(H_\nu)_0=\mathrm{Beta}(\nu+\frac12,1)$.
\item Given $(\lambda,d,\eta)$, $\theta$ is normal with precision $P_{\lambda,\eta}=\lambda_eW^TW+\diag(\Sigma_\beta^{-1},\Lambda_{\lambda,\eta})$ and mean $P_{\lambda,\eta}^{-1}\{\lambda_eW^Ty+((\Sigma_\beta^{-1}\mu_\beta)^T,0^T)^T\}$, a law that does not depend on $d$.
\end{enumerate}
Let $T_\lambda=\lambda_eX^TX+\Sigma_\beta^{-1}$, $M_\lambda=I_N-\lambda_eXT_\lambda^{-1}X^T$ and $Q_{\lambda,\eta}=\lambda_eZ^TM_\lambda Z+\Lambda_{\lambda,\eta}$. Taking the Schur complement of $T_\lambda$ in $P_{\lambda,\eta}$ gives $\Var(u\mid\lambda,\eta)=Q_{\lambda,\eta}^{-1}$.

\begin{algorithm}[t]
\caption{Block Gibbs sampler for model \eqref{eq:model-a}--\eqref{eq:model-c}.}\label{alg:gibbs}
\begin{algorithmic}[1]
\Require current state $(\eta,\theta)$ with $\theta=(\beta,u)$
\State Draw $\lambda_e$ and $\lambda_1,\dots,\lambda_K$ independently from their gamma full conditionals (i).
\State For $k=1,\dots,K$: draw $d_k$ from \eqref{eq:dpost}, then $\eta_i\sim(H_{i,d_k})_{s_i}$ for $i$ with $k(i)=k$.
\State Draw $\theta\sim\N_{p+q}\big(\E(\theta\mid\lambda,\eta),P_{\lambda,\eta}^{-1}\big)$.
\end{algorithmic}
\end{algorithm}

\subsection{The Markov chains}\label{sec:chains}
Algorithm~\ref{alg:gibbs} simulates a Markov chain $\Phi=\{(\lambda^{(n)},d^{(n)},\eta^{(n)},\theta^{(n)})\}_{n\ge0}$ whose invariant distribution is the posterior. The first step does not use $d$ or the previous $\lambda$, the second uses only $(\lambda,\theta)$, and the third uses only $(\lambda,\eta)$. Hence $\{(\eta^{(n)},\theta^{(n)})\}_{n\ge0}$ is itself a Markov chain, which we call the \emph{marginal chain}. Its state space is $\mathsf X=\prod_{i=1}^q\mathsf S_i\times\R^{p+q}$, where $\mathsf S_i$ is defined in assumption \ref{A0} below. We study the marginal chain and transfer the conclusions to $\Phi$ by Lemma~\ref{lem:topology}.

\section{Main results}\label{sec:results}

\subsection{Notation and standing assumption}
For a probability $H$ on $\R_+$, $r\in\R$ and $c\in(0,1]$, write $m_r(H)=\int\eta^rH(d\eta)\in(0,\infty]$ and, with $\E_s$ the expectation under $H_s$ in \eqref{eq:tilted},
\[
\kappa_H(c)=\limsup_{s\to\infty}\frac{\E_s[\eta^{-c}]}{s^c},\qquad R_H(c)=\frac{\Gamma(c+\frac12)}{\Gamma(\frac12)}\,\kappa_H(c).
\]
These are the quantities of \citet{delacruz:humeniy:2026}. For the Student-$t$ family with $d$ degrees of freedom and $c<(d+1)/2$, $(H_d)_s=\Ga(\frac{d+1}2,\frac d2+s)$, $\E_s[\eta^{-c}]=\Gamma(\frac{d+1}2-c)\Gamma(\frac{d+1}2)^{-1}(\frac d2+s)^c$ and
\[
R_{H_d}(c)=\frac{\Gamma(c+\frac12)\,\Gamma(\frac{d+1}2-c)}{\Gamma(\frac12)\,\Gamma(\frac{d+1}2)} .
\]
For the slash family with parameter $\nu$, $R_{H_\nu}(c)$ is given by the same formula with $(d+1)/2$ replaced by $\nu+\frac12$, for $c<\nu+\frac12$ \citep[Proposition~2(c)]{delacruz:humeniy:2026}. Both are decreasing in the tail parameter. For each coordinate $i$ let
\[
R_i(c)=\max_{d\in\mathcal D_{k(i)}}R_{H_{i,d}}(c).
\]

For the design, let $r=\rank(Z)$, let $Z_{-k}$ denote $Z$ without the block $Z_k$, $r_{-k}=\rank(Z_{-k})$ (with $r_{-k}=0$ if $K=1$), and let $\Pi_Z$ be the orthogonal projection of $\R^q$ onto $\mathcal N(Z)=\{v:Zv=0\}$. Define
\[
\ell_i=(\Pi_Z)_{ii},\qquad s_j(c)=\sum_{i=1}^q|(\Pi_Z)_{ij}|^{2c},\qquad S_k(c)=\max_{j:k(j)=k}s_j(c),\qquad \delta_k=q_k+r_{-k}-r .
\]
Thus $0\le\ell_i\le1$, $\sum_i\ell_i=q-r$, and $s_j(1)=\ell_j$, so that $S_k(1)=\ell_k^*:=\max_{i:k(i)=k}\ell_i$. The number $\delta_k\ge0$ is the dimension of the projection of $\mathcal N(Z)$ onto the coordinates of block $k$ (Lemma~\ref{lem:blocklev}), so $\delta_k=0$ exactly when the columns of $\Pi_Z$ for block $k$ vanish. We also write $\tau_Z=\tr\{(Z^TZ)^+\}$, where $A^+$ denotes the Moore--Penrose inverse.

We assume throughout:
\begin{enumerate}[label=\textup{(A0)},leftmargin=*]
\item\label{A0} For each $i$, the laws $H_{i,d}$, $d\in\mathcal D_{k(i)}$, are mutually absolutely continuous and satisfy $m_{1/2}(H_{i,d})<\infty$. We write $\mathsf S_i$ for their common support.
\end{enumerate}
For the Student-$t$ family on any grid, $\mathsf S_i=\R_+$; for the slash family, $\mathsf S_i=(0,1]$. A Harris ergodic Markov chain with kernel $P$ and invariant distribution $\pi$ is geometrically ergodic if there are $\rho<1$ and a function $C$ with $\norm{P^n(x,\cdot)-\pi}_{\mathrm{TV}}\le C(x)\rho^n$ for all $x$ and $n$.

\subsection{Geometric ergodicity}

\begin{theorem}\label{thm:main}
Assume \ref{A0} and
\begin{enumerate}[label=\textup{(C\arabic*)},leftmargin=*]
\item\label{C2} $a_e>\frac12(r-N+2)$;
\item\label{C3} for every $k$ with $\delta_k>0$, $a_k>1-\frac12(r-r_{-k})$.
\end{enumerate}
Suppose that there is $c\in(0,1]$ such that
\begin{enumerate}[label=\textup{(H\arabic*)},leftmargin=*]
\item\label{H1} $m_{c+1/2}(H_{i,d})<\infty$ and $\kappa_{H_{i,d}}(c)<\infty$ for every $i$ and every $d\in\mathcal D_{k(i)}$;
\item\label{H2} $R_j(c)\,s_j(c)<1$ for every $j=1,\dots,q$.
\end{enumerate}
Then the Gibbs chain $\Phi$ of Algorithm~\ref{alg:gibbs} is geometrically ergodic.
\end{theorem}

When the mixing laws depend only on the component, so that $R_i(c)=R_k(c)$ for $k(i)=k$, \ref{H2} reads $R_k(c)S_k(c)<1$ for every $k$, and it only restricts the components with $\delta_k>0$. When $Z$ has full column rank, \ref{C3} and \ref{H2} are void, and the conditions on the mixing laws reduce to \ref{H1}. For normal random effects (point mass at one), $\kappa_{H_{i,d}}(c)=0$. Condition \ref{C2} then coincides with the corresponding condition of \citet{roma:hobe:2014} for the block Gibbs sampler with proper priors, without their requirement that $X$ have full column rank, and \ref{C3} is no stronger, and potentially strictly weaker, than theirs when $K>1$, since $\delta_k\le q-r$ and \ref{C3} is imposed only on components with $\delta_k>0$. For $K=1$ and a known tail parameter, Theorem~\ref{thm:main} reduces to Theorem~1 of \citet{delacruz:humeniy:2026}. With $c=1$, \ref{H2} reads $\ell_jR_j(1)<1$, which for Student-$t$ random effects with grid minimum $d_{\min}>1$ is $d_{\min}>1+\ell_j$. Smaller values of $c$ weaken the moment condition in \ref{H1}. For the Student-$t$ and slash families $R_H(c)\to1$ as $c\downarrow0$, which helps when $R_H(1)>1$, that is, for $d<2$ or $\nu<1$; on the other hand $s_j(c)$ increases as $c$ decreases, towards the number of non-zero entries of column $j$ of $\Pi_Z$.

\subsection{Families with unknown tail parameters}

\begin{proposition}\label{prop:families}
Suppose that component $k$ has Student-$t$ or slash random effects with tail parameter on the finite grid $\mathcal D_k$, that is, $H_{i,d}=\Ga(d/2,d/2)$ or $H_{i,d}=\mathrm{Beta}(d,1)$ for $k(i)=k$, and let $d_{\min}=\min\mathcal D_k$. Then \ref{A0} holds for these coordinates, all positive moments of $H_{i,d}$ are finite, and $R_i(c)=R_{H_{d_{\min}}}(c)$ for $c<(d_{\min}+1)/2$ (Student-$t$) or $c<d_{\min}+\frac12$ (slash). In particular:
\begin{enumerate}[label=\textup{(\alph*)},leftmargin=*]
\item with $c=1$, \ref{H1}--\ref{H2} hold for these coordinates if $d_{\min}>1+\ell_k^*$ (Student-$t$) or $d_{\min}>(1+\ell_k^*)/2$ (slash);
\item if $\delta_k=0$, \ref{H1}--\ref{H2} hold for these coordinates for every $c\in(0,1]$ with $c<(d_{\min}+1)/2$ or $c<d_{\min}+\frac12$, so every finite grid is allowed;
\item if each column of $\Pi_Z$ for block $k$ is zero or a coordinate vector, so that $S_k\le1$ (zero columns of $Z$), \ref{H2} holds for these coordinates for small $c$ and every finite grid.
\end{enumerate}
For contaminated normal, normal-GIG with $\chi,\psi>0$, and variance-gamma random effects with shape $\alpha>\frac12$, with known parameters, $\kappa_H(c)=0$ for every $c$, so \ref{H2} holds for these coordinates, and \ref{H1} holds for every $c\in(0,1]$, except that $c<\alpha-\frac12$ is needed for the variance-gamma law \citep[Proposition~2 and Corollary~1]{delacruz:humeniy:2026}.
\end{proposition}

Since \ref{H1}--\ref{H2} must hold with the same $c$ for all components, in practice one evaluates $\max_jR_j(c)s_j(c)$ on a grid of values of $c$; Section~\ref{sec:app} gives an example.

\subsection{Rescaling the design}\label{sec:scaling}

Let $C=\diag(c_1,\dots,c_q)$ with $c_i>0$, and let $s_j(c;C)$ be computed as $s_j(c)$ with $\Pi_Z$ replaced by the orthogonal projection of $\R^q$ onto $C\,\mathcal N(Z)$. This is the projection onto the null space of $ZC^{-1}$, the design in which column $i$ of $Z$ has been divided by $c_i$.

\begin{corollary}\label{cor:scaling}
The conclusion of Theorem~\ref{thm:main} remains true if \ref{H2} is replaced by: there is a positive diagonal matrix $C$ such that $R_j(c)s_j(c;C)<1$ for every $j$.
\end{corollary}

The corollary is a consequence of an equivariance property. The rescaled model with design $ZC^{-1}$, random effects $Cu$ and weights $C^{-2}\eta$ is an equivalent parametrization of the same model, its Gibbs sampler is the image of Algorithm~\ref{alg:gibbs} under a fixed bijection, and the quantities in \ref{A0}, \ref{C2}--\ref{C3} and \ref{H1}, as well as $R_j(c)$, are unchanged by the rescaling (Section~\ref{sec:proof-scaling}). The projection onto the null space, however, does change.

\begin{example}[Proportional columns]\label{ex:pairs}
Suppose that $\mathcal N(Z)$ is spanned by vectors $w_1,\dots,w_m$ with disjoint supports, each supported on two coordinates, $w_j=(\dots,a_j,\dots,-b_j,\dots)$ with $a_j,b_j\ne0$. This is the situation of a random intercept and slope when the only linear dependencies come from subjects with a single observation at a time $t\ne0$: the intercept and slope columns of such a subject are $e$ and $te$, and $\mathcal N(Z)$ contains $(t,-1)$ on those two coordinates. On each pair, $\Pi_Z$ is the $2\times2$ matrix with entries $\ell$, $1-\ell$ on the diagonal and $\pm\{\ell(1-\ell)\}^{1/2}$ off the diagonal, where $\ell=a_j^2/(a_j^2+b_j^2)$; for the intercept of a subject observed once at time $t$, $\ell=t^2/(1+t^2)$. Taking $c_i=|b_j/a_j|$ for the first coordinate of each pair and $c_i=1$ elsewhere makes $\ell=\frac12$ for every pair, so that $s_j(c;C)=2^{1-2c}$ on the coordinates of the pairs and $s_j(\frac12;C)=1$. For the Student-$t$ family, $R_{H_d}(\frac12)=\Gamma(\frac d2)/\{\Gamma(\frac12)\Gamma(\frac{d+1}2)\}$, which is smaller than one exactly when $d>1$; for the slash family, $R_{H_\nu}(\frac12)<1$ exactly when $\nu>\frac12$. Hence, for Student-$t$ random intercepts and slopes, Corollary~\ref{cor:scaling} with $c=\frac12$ gives geometric ergodicity whenever all grids satisfy $\min\mathcal D_k>1$, whatever the number of subjects with a single observation, provided none of them is observed only at time zero. A subject observed once exactly at the origin has a zero slope column, so that $e_i\in\mathcal N(Z)$ and $s_i(c;C)=1$ for every $C$; for that coordinate \ref{H2} again reads $R_i(c)<1$, which holds at $c=\frac12$ under the same grid conditions.
\end{example}

Rescaling can be pushed further. Call a set $J$ of $q-r$ coordinates a \emph{pivot set} if the columns of $Z$ outside $J$ are linearly independent; such a set always exists, since the complement of any set of $r$ linearly independent columns is one.

\begin{corollary}\label{cor:anydesign}
Let $J$ be a pivot set. The conclusion of Theorem~\ref{thm:main} remains true if \ref{H2} is replaced by: $R_j(c)<1$ for every $j\in J$.
\end{corollary}

The idea is to multiply the coordinates in $J$ by a large constant: the null space of the rescaled design then concentrates on the coordinates of $J$, its projection approaches the diagonal matrix with ones on $J$, and the rescaled columns behave like zero columns of $Z$ (Section~\ref{sec:proof-anydesign}). For $K=1$ and known tail parameters this is Theorem~2 of \citet{delacruz:humeniy:2026}. Combined with Proposition~\ref{prop:families}, Corollary~\ref{cor:anydesign} shows that the rank of $Z$ imposes no restriction on the usual families.

\begin{corollary}\label{cor:families}
Assume \ref{A0} and \ref{C2}--\ref{C3}, and suppose that each component has Student-$t$ or slash random effects with tail parameters on finite grids, or contaminated normal, normal-GIG with $\chi,\psi>0$, or variance-gamma random effects with shape larger than $\frac12$ and known parameters. Then $\Phi$ is geometrically ergodic, whatever the design $Z$.
\end{corollary}

Theorem~\ref{thm:main} and Corollary~\ref{cor:scaling} remain useful when a family has $R_H(c)\ge1$ for all small $c$, and they show how the projection onto the null space enters the drift; for the families above, Corollary~\ref{cor:families} makes the condition on the tails disappear.

\subsection{Shared weights: multivariate scale mixtures}\label{sec:shared}

In random-coefficient models the coordinates of a subject's random-effect vector often share one latent weight, which makes the vector a multivariate scale mixture of normals, for instance multivariate Student-$t$ \citep{Pinheiro2001}; in that model the weight is also shared by the errors of the subject, whereas here the errors are normal. We now allow shared weights for the random effects. Let the coordinates $1,\dots,q$ be partitioned into blocks $b\in\mathcal B$, each contained in a single component, write $b(i)$ for the block of coordinate $i$, $n_b$ for the size of block $b$, $k(b)$ for its component and $u_b$ for the corresponding sub-vector of $u$. Model \eqref{eq:model-b} is replaced by
\begin{equation}\label{eq:model-block}
u_i\mid\lambda,\eta,d\ind\N\big(0,(\lambda_{k(i)}\eta_{b(i)})^{-1}\big),\qquad\eta_b\mid d\ind H_{b,d_{k(b)}},\qquad b\in\mathcal B,
\end{equation}
so that $u_b$ is a scale mixture of $\N_{n_b}(0,\lambda_{k(b)}^{-1}I)$ laws. Model \eqref{eq:model-b} is the case of blocks of size one. The full conditionals (i) and (iii) are unchanged, with $\eta_{b(i)}$ in place of $\eta_i$. In (ii), with $s_b=\lambda_{k(b)}\norm{u_b}^2/2$, the weight $\eta_b$ has the tilted law $(H_{b,d_{k(b)}})^{(n_b)}_{s_b}$, where, for a probability $H$ on $\R_+$ and $n\ge1$,
\[
\mathcal Z^{(n)}_H(s)=\int_0^\infty\eta^{n/2}e^{-s\eta}H(d\eta),\qquad H^{(n)}_s(d\eta)=\frac{\eta^{n/2}e^{-s\eta}H(d\eta)}{\mathcal Z^{(n)}_H(s)},
\]
and $\mathcal Z^{(n_b)}_{H_{b,d}}(s_b)$ replaces $\mathcal Z_{H_{i,d}}(s_i)$ in \eqref{eq:dpost}. For $H_d=\Ga(d/2,d/2)$, $(H_d)^{(n)}_s=\Ga\big(\frac{d+n}2,\frac d2+s\big)$. Let $\kappa^{(n)}_H(c)=\limsup_{s\to\infty}s^{-c}\E^{(n)}_s[\eta^{-c}]$, with $\E_s^{(n)}$ the expectation under $H^{(n)}_s$,
\[
R^{(n)}_H(c)=\frac{\Gamma(c+\frac n2)}{\Gamma(\frac n2)}\,\kappa^{(n)}_H(c),\qquad R_b(c)=\max_{d\in\mathcal D_{k(b)}}R^{(n_b)}_{H_{b,d}}(c),
\]
so that $R^{(1)}_H=R_H$. For blocks $b,b'$ let $F_{bb'}=\sum_{i\in b,\,j\in b'}(\Pi_Z)_{ij}^2$ be the squared Frobenius norm of the corresponding block of $\Pi_Z$, and
\[
\sigma_{b'}(c)=\sum_{b\in\mathcal B}\frac{\vartheta_{n_b}(c)}{\vartheta_{n_{b'}}(c)}\Big(\frac{F_{bb'}}{n_{b'}}\Big)^{c},\qquad \vartheta_n(c)=\frac{\Gamma(c+\frac n2)}{\Gamma(\frac n2)(\frac n2)^c}.
\]
When all blocks have the same size $n$, $\sigma_{b'}(c)=\sum_b(F_{bb'}/n)^c$; for $n=1$ it is $s_{b'}(c)$, and $\sigma_{b'}(1)=n_{b'}^{-1}\sum_{j\in b'}\ell_j$ in general.

\begin{theorem}\label{thm:block}
Consider model \eqref{eq:model-a}, \eqref{eq:model-block}, \eqref{eq:model-c}, and assume \ref{C2}--\ref{C3} and the block version of \ref{A0}: for each $b$, the laws $H_{b,d}$, $d\in\mathcal D_{k(b)}$, are mutually absolutely continuous, with common support $\mathsf S_b$, and $m_{n_b/2}(H_{b,d})<\infty$. The marginal chain then lives on $\prod_b\mathsf S_b\times\R^{p+q}$. Suppose that there is $c\in(0,1]$ such that $m_{c+n_b/2}(H_{b,d})<\infty$ and $\kappa^{(n_b)}_{H_{b,d}}(c)<\infty$ for every $b$ and $d\in\mathcal D_{k(b)}$, and $R_b(c)\,\sigma_b(c)<1$ for every $b\in\mathcal B$. Then the Gibbs chain is geometrically ergodic.
\end{theorem}

The condition involves $\Gamma(c+\frac n2)/\Gamma(\frac n2)$ instead of $\Gamma(c+\frac12)/\Gamma(\frac12)$, because the fractional moment of the norm of a block is that of a $\chi^2_{n_b}$ variable, and the tilted law gains the factor $\eta^{n_b/2}$. The most common case is a subject-specific design. Suppose that $m$ subjects have random-effect vectors $v_b\in\R^{n_0}$ with design matrices $Z^{[b]}$ ($N\times n_0$) whose non-zero rows are those of subject $b$, so that $Zu=\sum_bZ^{[b]}v_b$, and that
\begin{equation}\label{eq:mvt}
v_b\mid\lambda,\eta,d\ind\N_{n_0}\big(0,(\lambda_{g(b)}\eta_b\Psi_{g(b)})^{-1}\big),\qquad\eta_b\mid d\ind H_{g(b),d_{g(b)}},
\end{equation}
where the subjects are divided into groups $g$, each with a scale $\lambda_g$, a known positive definite shape matrix $\Psi_g$ and a tail parameter $d_g$ on a finite grid. For $H_d=\Ga(d/2,d/2)$, $v_b$ is multivariate Student-$t$ with $d_g$ degrees of freedom and scale matrix $(\lambda_g\Psi_g)^{-1}$. Let $\mathrm{nul}_b=n_0-\rank(Z^{[b]})$.

\begin{corollary}\label{cor:mvt}
Consider model \eqref{eq:model-a}, \eqref{eq:mvt}, \eqref{eq:model-c}, with the groups as components, and assume \ref{A0} with $m_{n_0/2}(H_{g,d})<\infty$ in place of $m_{1/2}$, and \ref{C2}--\ref{C3}. Suppose that there is $c\in(0,1]$ with $m_{c+n_0/2}(H_{g,d})<\infty$, $\kappa^{(n_0)}_{H_{g,d}}(c)<\infty$ for every group $g$ and $d\in\mathcal D_g$, and $R^{(n_0)}_{H_{g(b),d}}(c)\,(\mathrm{nul}_b/n_0)^c<1$ for every subject $b$ with $\mathrm{nul}_b>0$ and every $d\in\mathcal D_{g(b)}$. Then the Gibbs chain is geometrically ergodic. In particular this holds for multivariate Student-$t$ and slash random effects with any finite grids of tail parameters and any subject-specific design.
\end{corollary}

For multivariate Student-$t$ random effects, $\kappa^{(n_0)}_{H_d}(c)=\Gamma(\frac{d+n_0}2-c)/\Gamma(\frac{d+n_0}2)$ for $c<\frac{d+n_0}2$, and for the slash family the same holds with $\frac{d+n_0}2$ replaced by $\nu+\frac{n_0}2$. No rescaling is needed: for a subject-specific design the null space of $Z$ is the direct sum of the null spaces of the $Z^{[b]}$, so $\Pi_Z$ is block diagonal. The shape matrices $\Psi_g$ are known in \eqref{eq:mvt}; the case of an unknown, unstructured shape is discussed in Section~\ref{sec:disc}.

\subsection{Posterior moments}

Geometric ergodicity yields a Markov chain CLT for every function with a posterior moment of order larger than two, and consistent batch-means estimators of the asymptotic covariance under slightly stronger moment conditions \citep{Jones2004,vats:fleg:jone:2019,bane:vats:2024}. The next result supplies these moments for the quantities usually monitored.

\begin{proposition}\label{prop:moments}
In model \eqref{eq:model-a}--\eqref{eq:model-c}, and in the models with shared weights \eqref{eq:model-block} and \eqref{eq:mvt} of Section~\ref{sec:shared}, for all mixing laws, under the posterior distribution $\E\norm\beta^m<\infty$, $\E|\log\lambda_e|^m<\infty$ and $\E|\log\lambda_k|^m<\infty$ for every $m>0$ and every $k$.
\end{proposition}

The tail parameters take values in finite sets, so all their moments are finite, and the CLT applies to the indicators $\mathbf 1\{d_k=d\}$, which are bounded. Under the assumptions of Theorem~\ref{thm:main}, Corollary~\ref{cor:families}, Theorem~\ref{thm:block} or Corollary~\ref{cor:mvt}, the CLT therefore holds for $g=(\beta^T,\log\sigma_e,\log\sigma_1,\dots,\log\sigma_K)^T$, with $\sigma_e=\lambda_e^{-1/2}$ and $\sigma_k=\lambda_k^{-1/2}$, and for the posterior probabilities of the tail parameters. Since $g$ has posterior moments of all orders, the multivariate batch-means estimator with batch size $b_n=\lfloor n^{t_b}\rfloor$ is strongly consistent for every $t_b\in(\frac12,1)$. Indeed, for a geometrically ergodic chain and a function with posterior moments of order larger than four, Theorem~2(b) of \citet{bane:vats:2024} gives a strong invariance principle with rate $n^{1/4}\log n$, and the strong-consistency criterion of \citet{vats:fleg:jone:2019}, as restated in Theorem~3 of \citet{bane:vats:2024}, applies because $b_n^{-1}\log n\,(n^{1/4}\log n)^2=O\{n^{1/2-t_b}(\log n)^3\}$ tends to zero, that is, for $t_b>\frac12$.

\section{Proofs}\label{sec:proofs}

Norms of vectors are Euclidean and $\norm{A}_2$ is the spectral norm. For symmetric matrices, $A\pe B$ means that $B-A$ is non-negative definite.

\subsection{Results from the single-component case}
The following three results are Lemmas~2, 5 and~4 of \citet{delacruz:humeniy:2026}; to keep the paper self-contained, their proofs are given in Appendix~\ref{app:lemmas}. The first separates the row space of a design from its null space.

\begin{lemma}[\citealp{delacruz:humeniy:2026}, Lemma~2]\label{lem:trace}
Let $A$ be an $m\times n$ matrix, $\upsilon>0$, $\Upsilon=\diag(\upsilon_1,\dots,\upsilon_n)$ with $\upsilon_i>0$, and $\Omega=\upsilon A^TA+\Upsilon$. Let $\Pi_A$ be the orthogonal projection onto $\mathcal N(A)$, $P_A=I-\Pi_A$, and $\tilde\ell_i=(\Pi_A)_{ii}$. Then
\begin{enumerate}[label=\textup{(\alph*)},leftmargin=*]
\item $P_AQ^{-1}P_A\pe\upsilon^{-1}(A^TA)^+$ and $\Pi_AQ^{-1}\Pi_A\pe\Pi_A\Upsilon^{-1}\Pi_A$ for every $Q\succeq\Omega$;
\item $\tr(\Omega^{-1})\le\upsilon^{-1}\tr\{(A^TA)^+\}+\sum_{i=1}^n\tilde\ell_i\upsilon_i^{-1}$;
\item $\tr(A\Omega^{-1}A^T)\le\upsilon^{-1}\rank(A)$.
\end{enumerate}
\end{lemma}

The second concerns the tilted family \eqref{eq:tilted}.

\begin{lemma}[\citealp{delacruz:humeniy:2026}, Lemma~5]\label{lem:tilted}
Let $H$ be a probability on $\R_+$ with $m_{1/2}(H)<\infty$, and $c\in(0,1]$.
\begin{enumerate}[label=\textup{(\alph*)},leftmargin=*]
\item $\sup_{s\ge0}\E_s[\eta^c]=m_{c+1/2}(H)/m_{1/2}(H)$.
\item If $\kappa_H(c)<\infty$, then for every $\kappa'>\kappa_H(c)$ there is $C<\infty$ with $\E_s[\eta^{-c}]\le\kappa's^c+C$ for all $s\ge0$.
\end{enumerate}
\end{lemma}

The third bounds the conditional mean of $\theta$ uniformly in the weights. Its proof writes $\E(\theta\mid\lambda,\eta)$ as a weighted least squares solution and uses the bound of \citet{Stewart1989} and \citet{OLeary1990} on scaled projections. It uses only that the prior precision of $u$ is a positive diagonal matrix, so it applies verbatim with $\Lambda_{\lambda,\eta}$.

\begin{lemma}[\citealp{delacruz:humeniy:2026}, Lemma~4]\label{lem:means}
There are finite constants $K_1$ and $K_2$, depending only on $X$, $Z$, $y$, $\mu_\beta$ and $\Sigma_\beta$, such that $\norm{\E(\theta\mid\lambda,\eta)}\le K_1$ and $\norm{y-W\E(\theta\mid\lambda,\eta)}\le K_2$ for all $\lambda\in\R_+^{K+1}$ and $\eta\in\R_+^q$.
\end{lemma}

Finally, we use the following elementary facts \citep[Lemma~6]{delacruz:humeniy:2026}, also proved in Appendix~\ref{app:lemmas}. For $p>0$ and $\epsilon>0$ there is $C_{\epsilon,p}$ with $(x+y)^p\le(1+\epsilon)x^p+C_{\epsilon,p}y^p$ for $x,y\ge0$. If $Y\sim\N(m,\sigma^2)$ and $c\in(0,1]$, then $\E|Y|^{2c}\le(1+\epsilon)\mu_c\sigma^{2c}+C_{\epsilon,2c}|m|^{2c}$, where $\mu_c=\E|\xi|^{2c}=2^c\Gamma(c+\frac12)/\Gamma(\frac12)$ for $\xi\sim\N(0,1)$. We also use $(\sum_jw_j)^c\le\sum_jw_j^c$ for $w_j\ge0$, and Jensen's inequality $\E[X^c]\le(\E X)^c$ for $X\ge0$.
\subsection{Topological properties}

Let $P_m$ be the kernel of the marginal chain, $P_\Phi$ that of $\Phi$, $\pi$ the posterior and $\pi_m$ its $(\eta,\theta)$-marginal. For each $i$ fix $d_i^\circ\in\mathcal D_{k(i)}$, let $\bar H_i=H_{i,d_i^\circ}$ and, by \ref{A0}, let $h_{i,d}=dH_{i,d}/d\bar H_i$, which is positive $\bar H_i$-almost everywhere. Let $\varphi=\bigotimes_i\bar H_i\otimes\mathrm{Leb}_{p+q}$.

\begin{lemma}\label{lem:topology}
Under \ref{A0}, the marginal chain and $\Phi$ are Harris ergodic, and every compact subset of $\mathsf X$ is petite for the marginal chain. If the marginal chain is geometrically ergodic, so is $\Phi$.
\end{lemma}

\begin{proof}
Each step of Algorithm~\ref{alg:gibbs} draws one block from its full conditional, so $\pi$ is invariant for $P_\Phi$ and $\pi_m$ is invariant for $P_m$. Starting from $(\eta',\theta')$, one step of the marginal chain draws $\lambda\sim\pi(\lambda\mid\eta',\theta')$, then $(d,\eta)$ from its joint full conditional given $(\lambda,\theta')$, and then $\theta\sim\pi(\theta\mid\lambda,\eta)$. Summing over $d$, the law of $\eta$ given $(\lambda,\theta')$ has density with respect to $\bigotimes_i\bar H_i$
\[
g(\eta\mid\lambda,\theta')=\prod_{k=1}^K\ \sum_{d\in\mathcal D_k}\pi(d_k=d\mid\lambda,\theta')\prod_{i:k(i)=k}\frac{h_{i,d}(\eta_i)\,\eta_i^{1/2}e^{-s_i'\eta_i}}{\mathcal Z_{H_{i,d}}(s_i')},\qquad s_i'=\tfrac12\lambda_{k(i)}u_i'^2,
\]
so $P_m$ has the density $k(\eta,\theta\mid\eta',\theta')=\int\pi(\theta\mid\lambda,\eta)\,g(\eta\mid\lambda,\theta')\,\pi(\lambda\mid\eta',\theta')\,d\lambda$ with respect to $\varphi$, which is positive $\varphi$-almost everywhere because every factor is. Hence $P_m(x,A)>0$ for every $x\in\mathsf X$ whenever $\varphi(A)>0$, and the chain is $\varphi$-irreducible. It is aperiodic: if $D_1,\dots,D_{\varpi}$ with $\varpi\ge2$ were a periodic cycle, then $\varphi(D_1)>0$, because $P_m(x,D_1)=1$ for $x\in D_\varpi$ and $P_m(x,\cdot)$ has a density with respect to $\varphi$; and $P_m(x,D_1)=0$ for $x\in D_1$, a contradiction. The posterior density with respect to $\mathrm{Leb}_{K+1}\otimes(\text{counting measure})\otimes\bigotimes_iH_{i,d_{k(i)}}\otimes\mathrm{Leb}_{p+q}$ is positive, and by \ref{A0} $\pi_m$ is equivalent to $\varphi$. Thus $P_m(x,\cdot)\ll\pi_m$ for every $x$, and Corollary~1 of \citet{tier:1994} gives Harris recurrence; with the invariant probability $\pi_m$ the chain is Harris ergodic. The same argument applies to $\Phi$, whose kernel has a positive density with respect to $\mathrm{Leb}_{K+1}\otimes(\text{counting measure})\otimes\varphi$.

For the Feller property, we show that $k$ is continuous in $(\eta',\theta')$ and dominated locally uniformly by an integrable function. The gamma densities $\pi(\lambda\mid\eta',\theta')$ depend continuously on $(\eta',\theta')$, with rates bounded below by $b_e$ and $b_k$ and bounded above on compact sets. The weights $\pi(d_k=d\mid\lambda,\theta')\in[0,1]$ are continuous in $(\lambda,\theta')$, since each $\mathcal Z_{H_{i,d}}$ is continuous and positive. Fix the destination $(\eta,\theta)$, with $h_{i,d}(\eta_i)<\infty$ for every $i$ and $d$, which holds for $\varphi$-almost every destination; the factors that depend only on $(\eta,\theta)$ are then constants. For the tilted factors, let $\delta>0$. Since $\eta_i\in\mathsf S_i$, the support of $H_{i,d}$, we have $\mathcal Z_{H_{i,d}}(s)\ge e^{-s(\eta_i+\delta)}\int_{(0,\eta_i+\delta)}t^{1/2}H_{i,d}(dt)$ with a positive integral, so each factor is at most $C_\delta(\eta_i)h_{i,d}(\eta_i)\eta_i^{1/2}e^{s_i'\delta}$, where $C_\delta(\eta_i)$ is a maximum over the finite set $\mathcal D_{k(i)}$ and does not depend on $(\lambda,\eta',\theta')$. On a compact neighbourhood of $(\eta',\theta')$ on which $|u_i'|\le\bar u$ for all $i$, choose $\delta$ with $q\delta\bar u^2<\min_kb_k$; the product over $i$ is then at most a constant, depending on the fixed $\eta$, times $\exp(\frac12\sum_{k=1}^K b_k\lambda_k)$. Write $\lambda_0=\lambda_e$, $b_0=b_e$, $a_0^\star=a_e+\frac N2$ and $a_k^\star=a_k+\frac{q_k}2$. The full conditional of $\lambda_k$ is $\Ga(a_k^\star,\beta_k')$ with a rate $\beta_k'\ge b_k$ that is bounded above by some $\bar\beta_k$ on the neighbourhood, so its density is at most $\bar\beta_k^{a_k^\star}\Gamma(a_k^\star)^{-1}\lambda_k^{a_k^\star-1}e^{-b_k\lambda_k}$; and $\pi(\theta\mid\lambda,\eta)\le(2\pi)^{-(p+q)/2}\det(P_{\lambda,\eta})^{1/2}\le C(1+\sum_{k=0}^K\lambda_k)^m$ for some $C$ and $m$ depending on the fixed destination, because $P_{\lambda,\eta}$ is affine in $\lambda$. Hence the integrand in $\lambda$ is dominated, uniformly on the neighbourhood, by
\[
C'\Big(1+\sum_{k=0}^K\lambda_k\Big)^m\prod_{k=0}^K\lambda_k^{a_k^\star-1}\exp\Big(-\frac12\sum_{k=0}^Kb_k\lambda_k\Big),
\]
which is integrable over $\R_+^{K+1}$ because every $a_k^\star$ is positive and the tails decay exponentially. By dominated convergence in $\lambda$, $k(\eta,\theta\mid\eta',\theta')$ is continuous in $(\eta',\theta')$ for $\varphi$-almost every $(\eta,\theta)$, and by Scheff\'e's lemma $(\eta',\theta')\mapsto P_m((\eta',\theta'),\cdot)$ is continuous in total variation, so the chain is weak Feller. Since $\mathsf X$ is a closed subset of $\R_+^q\times\R^{p+q}$, it is locally compact and separable, and $\mathrm{supp}(\varphi)=\mathsf X$. Theorems~6.2.9 and~6.2.5 of \citet{meyn:twee:1993} show that the chain is a T-chain and that every compact set is petite.

Finally, $\Phi^{(n+1)}$ is obtained from $(\eta^{(n)},\theta^{(n)})$ by a fixed kernel $R$ with $\pi_mR=\pi$. Since applying a Markov kernel does not increase total variation distance, $\norm{P_\Phi^{n+1}(x,\cdot)-\pi}_{\mathrm{TV}}\le\norm{P_m^n((\eta,\theta),\cdot)-\pi_m}_{\mathrm{TV}}$ for $x=(\lambda,d,\eta,\theta)$ and $n\ge0$. If the right-hand side is at most $C_m(\eta,\theta)\rho^n$, then $\norm{P_\Phi^{n}(x,\cdot)-\pi}_{\mathrm{TV}}\le\max\{1,C_m(\eta,\theta)/\rho\}\rho^n$ for all $n\ge0$.
\end{proof}

By Lemma~\ref{lem:topology} and Lemma~15.2.8 and Theorem~15.0.1 of \citet{meyn:twee:1993}, Theorem~\ref{thm:main} follows once we find $v:\mathsf X\to[1,\infty)$ with compact sublevel sets such that, for some $\rho<1$ and $L<\infty$,
\begin{equation}\label{eq:drift}
\E\big[v(\eta,\theta)\mid\eta',\theta'\big]\le\rho\,v(\eta',\theta')+L\qquad\text{for all }(\eta',\theta')\in\mathsf X,
\end{equation}
where the expectation is with respect to $P_m((\eta',\theta'),\cdot)$.

\subsection{Block leverages}

\begin{lemma}\label{lem:blocklev}
Let $\Lambda$ be a positive diagonal $q\times q$ matrix, let $\mathcal S=\Lambda\,\mathcal N(Z)$ and let $\Pi_{\mathcal S}$ be the orthogonal projection onto $\mathcal S$. For every $k$, $\sum_{i:k(i)=k}(\Pi_{\mathcal S})_{ii}\le\delta_k$, and $\delta_k$ is the dimension of the projection of $\mathcal N(Z)$ onto the coordinates of block $k$.
\end{lemma}

\begin{proof}
Let $P_k$ be the orthogonal projection onto the coordinates of block $k$. Then $\sum_{i:k(i)=k}(\Pi_{\mathcal S})_{ii}=\tr(P_k\Pi_{\mathcal S}P_k)$. The matrix $P_k\Pi_{\mathcal S}P_k$ is non-negative definite with eigenvalues in $[0,1]$, so its trace is at most its rank, which is at most $\rank(P_k\Pi_{\mathcal S})=\dim P_k\mathcal S$. Since $\Lambda$ is diagonal and invertible, it commutes with $P_k$ and $\dim P_k\mathcal S=\dim P_k\Lambda\mathcal N(Z)=\dim\Lambda P_k\mathcal N(Z)=\dim P_k\mathcal N(Z)$. Finally, the kernel of the map $v\mapsto P_kv$ restricted to $\mathcal N(Z)$ consists of the $v\in\mathcal N(Z)$ that vanish on block $k$, which correspond to the elements of $\mathcal N(Z_{-k})$; hence $\dim P_k\mathcal N(Z)=\dim\mathcal N(Z)-\dim\mathcal N(Z_{-k})=(q-r)-(q-q_k-r_{-k})=\delta_k$.
\end{proof}

\subsection{Conditional moments}

Let $\varsigma_{\max}$ be the largest singular value of $X\Sigma_\beta^{1/2}$, $c_\lambda^{-1}=\lambda_e^{-1}+\varsigma_{\max}^2$ and $\Lambda_\lambda=\diag(\lambda_{k(1)},\dots,\lambda_{k(q)})$. As in \citet[Section~4.2]{delacruz:humeniy:2026}, $(1+\lambda_e\varsigma_{\max}^2)^{-1}I\pe M_\lambda\pe I$ and $XT_\lambda^{-1}X^T\pe X\Sigma_\beta X^T$, so $\lambda_eZ^TM_\lambda Z\succeq c_\lambda Z^TZ$ and $Q_{\lambda,\eta}\succeq c_\lambda Z^TZ+\Lambda_{\lambda,\eta}$.

\begin{lemma}\label{lem:traces}
For all $\lambda\in\R_+^{K+1}$ and $\eta\in\R_+^q$,
\begin{enumerate}[label=\textup{(\alph*)},leftmargin=*]
\item $\tr(\Deta^{1/2}Q_{\lambda,\eta}^{-1}\Deta^{1/2})\le c_\lambda^{-1}\tau_Z\max_i\eta_i+\sum_{k=1}^K\delta_k\lambda_k^{-1}$;
\item $\tr\{\Var(W\theta\mid\lambda,\eta)\}\le c_\lambda^{-1}r+\tr(X\Sigma_\beta X^T)$.
\end{enumerate}
\end{lemma}

\begin{proof}
(a) We have $\Deta^{1/2}Q_{\lambda,\eta}^{-1}\Deta^{1/2}=(\Deta^{-1/2}Q_{\lambda,\eta}\Deta^{-1/2})^{-1}\pe(c_\lambda\Deta^{-1/2}Z^TZ\Deta^{-1/2}+\Lambda_\lambda)^{-1}$. Apply Lemma~\ref{lem:trace}(b) with $A=Z\Deta^{-1/2}$, $\upsilon=c_\lambda$ and $\Upsilon=\Lambda_\lambda$. The null space of $A$ is $\Deta^{1/2}\mathcal N(Z)$, so by Lemma~\ref{lem:blocklev} with $\Lambda=\Deta^{1/2}$ the leverages $\tilde\ell_i$ of $A$ satisfy $\sum_{i:k(i)=k}\tilde\ell_i\le\delta_k$, and $\sum_i\tilde\ell_i\lambda_{k(i)}^{-1}\le\sum_k\delta_k\lambda_k^{-1}$. Moreover $\tr\{(A^TA)^+\}=\tr\{(AA^T)^+\}=\tr\{(Z\Deta^{-1}Z^T)^+\}$. The matrices $Z\Deta^{-1}Z^T$ and $ZZ^T$ have the same range and $Z\Deta^{-1}Z^T\succeq(\max_i\eta_i)^{-1}ZZ^T$; restricting both to their common range, where they are positive definite, and inverting gives $(Z\Deta^{-1}Z^T)^+\pe\max_i\eta_i(ZZ^T)^+$. Since $\tr\{(ZZ^T)^+\}=\tau_Z$, (a) follows.

(b) Given $(\lambda,\eta,u)$, $\beta$ is normal with precision $T_\lambda$, and $\E(X\beta+Zu\mid\lambda,\eta,u)$ equals a constant plus $M_\lambda Zu$. By the law of total variance, $\Var(W\theta\mid\lambda,\eta)=XT_\lambda^{-1}X^T+M_\lambda ZQ_{\lambda,\eta}^{-1}Z^TM_\lambda$. The trace of the first term is at most $\tr(X\Sigma_\beta X^T)$. Since $0\pe M_\lambda^2\pe I$, the trace of the second is at most $\tr(ZQ_{\lambda,\eta}^{-1}Z^T)\le\tr\{Z(c_\lambda Z^TZ+\Lambda_{\lambda,\eta})^{-1}Z^T\}\le c_\lambda^{-1}r$ by Lemma~\ref{lem:trace}(c).
\end{proof}

\subsection{Proof of Theorem~\ref{thm:main}}

Fix $c$ as in \ref{H1}--\ref{H2}, and let $\mathcal I=\{(i,d):1\le i\le q,\ d\in\mathcal D_{k(i)}\}$, a finite set. Since $m_{1/2}\le1+m_{c+1/2}$, \ref{A0} and Lemma~\ref{lem:tilted} apply to every $H_{i,d}$. Let $M_c=\max_{(i,d)\in\mathcal I}m_{c+1/2}(H_{i,d})/m_{1/2}(H_{i,d})<\infty$ and $\kappa_i(c)=\max_{d\in\mathcal D_{k(i)}}\kappa_{H_{i,d}}(c)$. Let $\epsilon>0$ and $\kappa_i'=\kappa_i(c)+\epsilon$. By \ref{H2} we may choose $\epsilon$ so small that
\[
\bar\gamma:=(1+\epsilon)^2\,\frac{\Gamma(c+\frac12)}{\Gamma(\frac12)}\,\max_j\kappa_j's_j(c)<1 .
\]
By Lemma~\ref{lem:tilted}, applied to each $H_{i,d}$ and taking the maximum of the constants over $\mathcal I$, there is $C_0<\infty$ such that $\E_{(H_{i,d})_s}[\eta^c]\le M_c$ and $\E_{(H_{i,d})_s}[\eta^{-c}]\le\kappa_i's^c+C_0$ for all $(i,d)\in\mathcal I$ and $s\ge0$.

Fix $(\eta',\theta')$ and write $\E'$ for expectation under one step of the marginal chain from $(\eta',\theta')$. Given $(\lambda,\theta')$, the law of $\eta_j$ is a mixture over $d_{k(j)}$ of the laws $(H_{j,d})_{s_j'}$, $s_j'=\lambda_{k(j)}u_j'^2/2$, and does not involve $\lambda_e$; hence
\begin{equation}\label{eq:etamoments}
\E[\eta_j^c\mid\lambda,\theta']\le M_c,\qquad
\lambda_{k(j)}^{-c}\E[\eta_j^{-c}\mid\lambda,\theta']\le\kappa_j'2^{-c}|u_j'|^{2c}+C_0\lambda_{k(j)}^{-c}.
\end{equation}
From the gamma full conditionals and Jensen's inequality,
\begin{equation}\label{eq:lambdamoments}
\E'[\lambda_e^{-c}]\le\Big(\frac{2b_e+\norm{y-W\theta'}^2}{2a_e+N-2}\Big)^c,\qquad
\E'[\lambda_k^{-c}]\le\Big(\frac{2b_k+\norm{D_{\eta',k}^{1/2}u_k'}^2}{2a_k+q_k-2}\Big)^c,\qquad
\E'[\lambda_k^c]\le\zeta_k^c,
\end{equation}
where $\zeta_k=(2a_k+q_k)/(2b_k)$, and $u_k$ and $D_{\eta,k}$ are the sub-vector and sub-matrix of $u$ and $\Deta$ for block $k$. The first bound is valid because $2a_e+N-2>r\ge0$ by \ref{C2}. The second is used only for $k$ with $\delta_k>0$, and is then valid because \ref{C3} and $r-r_{-k}\le q_k$ give $2a_k+q_k-2>q_k-(r-r_{-k})\ge0$. Also $c_\lambda^{-c}\le\lambda_e^{-c}+\varsigma_{\max}^{2c}$. Let
\[
v(\eta,\theta)=\alpha_1\sum_{i=1}^q|u_i|^{2c}+\alpha_2\norm{\Deta^{1/2}u}^{2c}+\alpha_3\norm{y-W\theta}^{2c}+\sum_{i=1}^q(\eta_i^c+\eta_i^{-c})+\norm{\beta}^{2c},
\]
with $\alpha_1,\alpha_3>0$ to be chosen, $\alpha_2>0$ to be chosen if some $\delta_k>0$, and $\alpha_2=0$ otherwise. Write $V_1,\dots,V_5$ for the five terms without their coefficients and $V_k'$ for their values at $(\eta',\theta')$. Let $\mathcal K_+=\{k:\delta_k>0\}$. Then $v\ge2q\ge1$, and constants $C_1,C_2,\dots$ and $L_1,L_2,\dots$ below do not depend on $(\eta',\theta')$.

\emph{Sublevel sets are compact.} Let $\bar v\ge1$. If $v(\eta,\theta)\le\bar v$, then each $\eta_i$ lies in $[\bar v^{-1/c},\bar v^{1/c}]$, $|u_i|^{2c}\le\bar v/\alpha_1$ for every $i$ and $\norm\beta^{2c}\le\bar v$. The set $\{v\le\bar v\}$ is closed in $\mathsf X$ because $v$ is continuous, and it is contained in $\prod_i([\bar v^{-1/c},\bar v^{1/c}]\cap\mathsf S_i)\times\bar B$ for a closed ball $\bar B$, which is closed and bounded in $\R^{2q+p}$ because each $\mathsf S_i$ is closed in $\R_+$. Hence $\{v\le\bar v\}$ is compact.

\emph{The term $V_1$.} Write $\Pi=\Pi_Z$ and $P=I-\Pi$. Given $(\lambda,\eta)$, $u_i\sim\N(m_i,\sigma_i^2)$ with $|m_i|\le K_1$ by Lemma~\ref{lem:means} and $\sigma_i=\norm{Q_{\lambda,\eta}^{-1/2}e_i}\le\norm{Q_{\lambda,\eta}^{-1/2}\Pi e_i}+\norm{Q_{\lambda,\eta}^{-1/2}Pe_i}$. Lemma~\ref{lem:trace}(a), with $A=Z$, $\upsilon=c_\lambda$, $\Upsilon=\Lambda_{\lambda,\eta}$ and $Q=Q_{\lambda,\eta}$, gives $e_i^T\Pi Q_{\lambda,\eta}^{-1}\Pi e_i\le\sum_j\Pi_{ij}^2(\lambda_{k(j)}\eta_j)^{-1}$ and $e_i^TPQ_{\lambda,\eta}^{-1}Pe_i\le c_\lambda^{-1}\{(Z^TZ)^+\}_{ii}$. Hence
\[
\sigma_i^{2c}\le(1+\epsilon)\sum_j|\Pi_{ij}|^{2c}\lambda_{k(j)}^{-c}\eta_j^{-c}+C_1c_\lambda^{-c},
\]
\[
\E(V_1\mid\lambda,\eta)\le(1+\epsilon)^2\mu_c\sum_j s_j(c)\lambda_{k(j)}^{-c}\eta_j^{-c}+C_2c_\lambda^{-c}+C_3 .
\]
A column $j$ of $\Pi$ is non-zero only if $\delta_{k(j)}>0$. By \eqref{eq:etamoments} and $\mu_c2^{-c}=\Gamma(c+\frac12)/\Gamma(\frac12)$,
\[
\E(V_1\mid\lambda,\theta')\le\bar\gamma V_1'+C_4\sum_{k:\delta_k>0}\lambda_k^{-c}+C_2c_\lambda^{-c}+C_3 .
\]
By \eqref{eq:lambdamoments}, $(x+y)^c\le x^c+y^c$, and $\norm{D_{\eta',k}^{1/2}u_k'}\le\norm{D_{\eta'}^{1/2}u'}$,
\[
\E'V_1\le\bar\gamma V_1'+A_{12}V_2'+A_{13}V_3'+L_1
\]
for finite constants $A_{12}$ (zero if $\mathcal K_+=\emptyset$) and $A_{13}$.

\emph{The term $V_2$ (only if some $\delta_k>0$).} By Lemmas~\ref{lem:traces}(a) and~\ref{lem:means}, $\max_i\eta_i\le\sum_i\eta_i$ and $\delta_k=0$ for $k\notin\mathcal K_+$,
\[
\E(\norm{\Deta^{1/2}u}^2\mid\lambda,\eta)\le\big(c_\lambda^{-1}\tau_Z+K_1^2\big)\sum_i\eta_i+\sum_{k\in\mathcal K_+}\delta_k\lambda_k^{-1},
\]
By Jensen's inequality and $(x+y)^c\le x^c+y^c$,
\[
\E(V_2\mid\lambda,\eta)\le\big(c_\lambda^{-c}\tau_Z^c+K_1^{2c}\big)\sum_i\eta_i^c+\Big(\sum_{k\in\mathcal K_+}\delta_k\lambda_k^{-1}\Big)^c .
\]
The first term has $\E'$-expectation at most $qM_c(\tau_Z^c\E'[c_\lambda^{-c}]+K_1^{2c})\le A_{23}V_3'+C_5$ by \eqref{eq:etamoments} and \eqref{eq:lambdamoments}. For the second, Jensen's inequality with respect to the law of $\lambda$ given $(\eta',\theta')$ and the formula for $\E'[\lambda_k^{-1}]$, which is finite for $k\in\mathcal K_+$ by \ref{C3}, give
\[
\E'\Big(\sum_{k\in\mathcal K_+}\delta_k\lambda_k^{-1}\Big)^c\le\Big(\sum_{k\in\mathcal K_+}\delta_k\frac{2b_k+\norm{D_{\eta',k}^{1/2}u_k'}^2}{2a_k+q_k-2}\Big)^c\le C_6+A_{22}V_2',
\]
with
\[
A_{22}=\Big(\max_{k\in\mathcal K_+}\frac{\delta_k}{2a_k+q_k-2}\Big)^c,
\]
because $\sum_{k\in\mathcal K_+}\norm{D_{\eta',k}^{1/2}u_k'}^2\le\norm{D_{\eta'}^{1/2}u'}^2$. For $k\in\mathcal K_+$, condition \ref{C3} is equivalent to $\delta_k<2a_k+q_k-2$, so $A_{22}<1$, and
\[
\E'V_2\le A_{22}V_2'+A_{23}V_3'+L_2 .
\]

\emph{The term $V_3$.} By Lemmas~\ref{lem:traces}(b) and~\ref{lem:means}, $\E(\norm{y-W\theta}^2\mid\lambda,\eta)\le c_\lambda^{-1}r+\tr(X\Sigma_\beta X^T)+K_2^2$. By Jensen's inequality, $(x+y)^c\le x^c+y^c$ and \eqref{eq:lambdamoments},
\[
\E'V_3\le r^c\E'[\lambda_e^{-c}]+C_7\le A_{33}V_3'+L_3,\qquad A_{33}=\Big(\frac{r}{2a_e+N-2}\Big)^c<1
\]
by \ref{C2}.

\emph{The term $V_4$.} By \eqref{eq:etamoments} and \eqref{eq:lambdamoments},
\[
\E'V_4\le qM_c+\sum_j\big(\kappa_j'2^{-c}\E'[\lambda_{k(j)}^c]\,|u_j'|^{2c}+C_0\big)\le\bar\kappa2^{-c}\bar\zeta^cV_1'+L_4,
\]
with $\bar\kappa=\max_j\kappa_j'$ and $\bar\zeta=\max_k\zeta_k$.

\emph{The term $V_5$.} Given $(\lambda,\eta)$, $\beta$ is normal with mean of norm at most $K_1$ by Lemma~\ref{lem:means}, and its covariance matrix is at most $\Sigma_\beta$ in the positive semidefinite order, because $P_{\lambda,\eta}\succeq\diag(\Sigma_\beta^{-1},\Lambda_{\lambda,\eta})$. The bound on $\E|Y|^{2c}$ for normal $Y$, applied to each coordinate, and $(\sum_jw_j)^c\le\sum_jw_j^c$ give $\E(V_5\mid\lambda,\eta)\le L_5$ for a constant $L_5$, so $\E'V_5\le L_5$. No condition on the rank of $X$ is needed.

\emph{Conclusion.} Collecting terms,
\[
\E'v\le\Big(\bar\gamma+\frac{\bar\kappa2^{-c}\bar\zeta^c}{\alpha_1}\Big)\alpha_1V_1'+\Big(A_{22}+\frac{\alpha_1A_{12}}{\alpha_2}\Big)\alpha_2V_2'+\Big(A_{33}+\frac{\alpha_1A_{13}+\alpha_2A_{23}}{\alpha_3}\Big)\alpha_3V_3'+L,
\]
where the second term is absent if $\mathcal K_+=\emptyset$. If $\mathcal K_+\ne\emptyset$, choose first $\alpha_1>\bar\kappa2^{-c}\bar\zeta^c/(1-\bar\gamma)$, then $\alpha_2>\alpha_1A_{12}/(1-A_{22})$, then $\alpha_3>(\alpha_1A_{13}+\alpha_2A_{23})/(1-A_{33})$. If $\mathcal K_+=\emptyset$, then $\alpha_2=0$, $A_{12}=0$, the term $V_2$ and the constants $A_{22}$ and $A_{23}$ do not appear, and we choose $\alpha_1>\bar\kappa2^{-c}\bar\zeta^c/(1-\bar\gamma)$ and then $\alpha_3>\alpha_1A_{13}/(1-A_{33})$. In both cases every coefficient is smaller than one, and since $V_4',V_5'\ge0$ and $\E'V_5\le L_5$, \eqref{eq:drift} holds with some $\rho<1$. \qed

\subsection{Proof of Proposition~\ref{prop:families}}

All the laws $\Ga(d/2,d/2)$, $d>0$, are equivalent to Lebesgue measure on $\R_+$ and have all positive moments, and all the laws $\mathrm{Beta}(\nu,1)$, $\nu>0$, are equivalent to Lebesgue measure on $(0,1)$; so \ref{A0} holds and $m_{c+1/2}<\infty$ for every $c$. The formula for $R_{H_d}(c)$ follows from the Gamma form of the tilted law, and that for the slash family from the asymptotics $\E_s[\eta^{-c}]\sim\{\Gamma(a+\frac12-c)/\Gamma(a+\frac12)\}s^c$ for laws with index $a$ at zero \citep[Proposition~2(c)]{delacruz:humeniy:2026}, with $a=\nu$. Since $x\mapsto\Gamma(x-c)/\Gamma(x)$ is decreasing on $(c,\infty)$ for $c>0$, $R_{H_d}(c)$ is decreasing in the tail parameter and the maximum over the grid is attained at $d_{\min}$. (a) With $c=1$, $R_{H_d}(1)=1/(d-1)$ for the Student-$t$ and $1/(2\nu-1)$ for the slash law, and $s_j(1)=\ell_j\le\ell_k^*$. (b) If $\delta_k=0$, the columns of $\Pi_Z$ for block $k$ vanish, so $s_j(c)=0$ for these $j$. (c) If $S_k\equiv1$, \ref{H2} reads $R_{H_{d_{\min}}}(c)<1$; since $R_{H}(0)=1$ and $\frac{d}{dc}\log R_H(c)$ at $c=0$ equals $\psi(\frac12)-\psi(a+\frac12)<0$, where $\psi$ is the digamma function and $a=d_{\min}/2$ or $a=d_{\min}$, this holds for small $c>0$. The last statement follows from Proposition~2(a),(b) and Corollary~1(iii)--(v) of \citet{delacruz:humeniy:2026}. \qed

\subsection{Proof of Proposition~\ref{prop:moments}}

The argument is that of \citet[Proposition~3]{delacruz:humeniy:2026} and applies verbatim to the three models, because it uses only the likelihood and the priors of $\beta$ and the precisions, which are common to them. For all $(\beta,u,\lambda_e)$, $p(y\mid\beta,u,\lambda_e)\le(\lambda_e/2\pi)^{N/2}$. Hence, for every non-negative function $F$ of the parameters,
\[
\E_\pi[F]\le\frac{(2\pi)^{-N/2}}{m(y)}\,\E_{\mathrm{prior}}\big[\lambda_e^{N/2}F\big],
\]
where $m(y)$ is the marginal likelihood. Under the prior, $\beta,\lambda_e,\lambda_1,\dots,\lambda_K$ are independent, $\beta$ is normal and the precisions are gamma with positive shapes and rates. Taking $F=\norm\beta^m$, $F=|\log\lambda_e|^m$ and $F=|\log\lambda_k|^m$, the right-hand side is finite because $\E[\lambda_e^{N/2}]$, $\E\norm\beta^m$, $\E[\lambda_e^{N/2}|\log\lambda_e|^m]$ and $\E|\log\lambda_k|^m$ are finite under these priors. The mixing laws and the tail parameters play no role. \qed

\subsection{Proof of Corollary~\ref{cor:scaling}}\label{sec:proof-scaling}

Let $\tilde Z=ZC^{-1}$, and for each $(i,d)$ let $\tilde H_{i,d}$ be the law of $\eta/c_i^2$ when $\eta\sim H_{i,d}$. Consider model \eqref{eq:model-a}--\eqref{eq:model-c} with $Z$ and $H_{i,d}$ replaced by $\tilde Z$ and $\tilde H_{i,d}$ and all priors unchanged, and the bijection
\[
\mathcal T(\lambda,d,\eta,\theta)=\big(\lambda,d,C^{-2}\eta,(\beta,Cu)\big).
\]
If $u_i\mid\lambda,\eta\sim\N(0,(\lambda_{k(i)}\eta_i)^{-1})$, then $c_iu_i\mid\lambda,\eta\sim\N(0,(\lambda_{k(i)}\eta_i/c_i^2)^{-1})$, and $\tilde Z(Cu)=Zu$. Hence the joint law of $(\mathcal T(\lambda,d,\eta,\theta),Y)$ under the original model is the joint law of the parameters and $Y$ under the rescaled model. Consequently the posterior of the rescaled model is the image of $\pi$ under $\mathcal T$. Since the chains must agree from every starting point, we check directly that each full conditional of the rescaled model, evaluated at $\mathcal T(x)$, is the image under the corresponding coordinate map of the full conditional of the original model at $x$. For $\lambda$, the gamma parameters depend on $\norm{y-W\theta}^2$ and on $\sum_{i:k(i)=k}\eta_iu_i^2$, which are unchanged. For $(d,\eta)$, the tilting parameter of coordinate $i$ becomes $\lambda_{k(i)}c_i^2u_i^2/2=c_i^2s_i$, and $\mathcal Z_{\tilde H_{i,d}}(c_i^2s_i)=\int(\eta/c_i^2)^{1/2}e^{-s_i\eta}H_{i,d}(d\eta)=c_i^{-1}\mathcal Z_{H_{i,d}}(s_i)$; the factor $\prod_{i:k(i)=k}c_i^{-1}$ does not depend on $d_k$ and cancels in \eqref{eq:dpost}, so the conditional law of $d_k$ is unchanged, and, given $d_k$, the tilted law $(\tilde H_{i,d})_{c_i^2s_i}$ is the law of $\eta/c_i^2$ with $\eta\sim(H_{i,d})_{s_i}$. For $\theta$, the conditional law is Gaussian and the map $\theta\mapsto(\beta,Cu)$ is linear, with $\tilde W(\beta,Cu)=W\theta$ and $\sum_i\lambda_{k(i)}(\eta_i/c_i^2)(c_iu_i)^2=\sum_i\lambda_{k(i)}\eta_iu_i^2$, so the conditional law of the rescaled model is the image of the original one. Therefore the Gibbs chain $\tilde\Phi$ of the rescaled model, started at $\mathcal T(x)$, has the law of $\mathcal T(\Phi)$ started at $x$, and $\norm{\tilde P^n(\mathcal T(x),\cdot)-\tilde\pi}_{\mathrm{TV}}=\norm{P_\Phi^n(x,\cdot)-\pi}_{\mathrm{TV}}$ for every $n$, because total variation distance is invariant under bijections that are measurable with measurable inverses. It thus suffices to verify the hypotheses of Theorem~\ref{thm:main} for the rescaled model.

Conditions \ref{C2}--\ref{C3} are unchanged, because $\rank(\tilde Z)=\rank(Z)$ and $\rank(\tilde Z_{-k})=\rank(Z_{-k})$. The laws $\tilde H_{i,d}$, $d\in\mathcal D_{k(i)}$, are mutually absolutely continuous with support $c_i^{-2}\mathsf S_i$, and $m_r(\tilde H_{i,d})=c_i^{-2r}m_r(H_{i,d})$, so \ref{A0} and the moment part of \ref{H1} hold. For the tilted laws, $\tilde H_{i,d}$ tilted at $s$ is the law of $\eta/c_i^2$ with $\eta\sim(H_{i,d})_{s/c_i^2}$, because $x^{1/2}e^{-sx}$ with $x=\eta/c_i^2$ is proportional to $\eta^{1/2}e^{-(s/c_i^2)\eta}$. Hence $\E_{(\tilde H_{i,d})_s}[x^{-c}]=c_i^{2c}\E_{(H_{i,d})_{s/c_i^2}}[\eta^{-c}]$ and
\[
\kappa_{\tilde H_{i,d}}(c)=\limsup_{s\to\infty}\frac{c_i^{2c}\E_{(H_{i,d})_{s/c_i^2}}[\eta^{-c}]}{s^c}=\limsup_{t\to\infty}\frac{\E_{(H_{i,d})_t}[\eta^{-c}]}{t^c}=\kappa_{H_{i,d}}(c) .
\]
So $R_j(c)$ is unchanged, while the projection onto $\mathcal N(\tilde Z)=C\mathcal N(Z)$ gives $s_j(c;C)$. Condition \ref{H2} for the rescaled model is the hypothesis of the corollary. \qed

\subsection{Proof of Corollaries~\ref{cor:anydesign} and~\ref{cor:families}}\label{sec:proof-anydesign}

Let $m=q-r$ and order the coordinates so that $J=\{1,\dots,m\}$. Since the columns of $Z$ outside $J$ are linearly independent and span the column space of $Z$, for every $v_J\in\R^m$ there is exactly one $v\in\mathcal N(Z)$ with these first $m$ coordinates. Hence $\mathcal N(Z)$ is the column space of a matrix $\begin{pmatrix}I_m\\A\end{pmatrix}$, with $A$ of order $(q-m)\times m$. For $M\ge1$ let $C_M=\diag(M1_m,1_{q-m})$. Then $C_M\mathcal N(Z)$ is the column space of $B_M=\begin{pmatrix}I_m\\A/M\end{pmatrix}$, and its orthogonal projection is $\Pi_M=B_M(B_M^TB_M)^{-1}B_M^T$. As $M\to\infty$, $B_M^TB_M=I_m+A^TA/M^2\to I_m$, so $\Pi_M\to\diag(1_m,0_{q-m})$ entrywise. For fixed $c\in(0,1]$, $x\mapsto|x|^{2c}$ is continuous, so $s_j(c;C_M)\to1$ for $j\in J$ and $s_j(c;C_M)\to0$ for $j\notin J$. If $R_j(c)<1$ for $j\in J$, then $R_j(c)s_j(c;C_M)<1$ for every $j$ once $M$ is large enough, and Corollary~\ref{cor:scaling} gives Corollary~\ref{cor:anydesign}.

For Corollary~\ref{cor:families}, take any pivot set $J$. For a Student-$t$ or slash component, Proposition~\ref{prop:families} gives \ref{H1} for every small $c$ and $R_j(c)=R_{H_{d_{\min}}}(c)<1$ for small $c>0$, by the derivative computation in its proof. For the other families, $\kappa_H(c)=0$, so $R_j(c)=0$, and \ref{H1} holds for every $c\in(0,1]$ for the contaminated normal and normal-GIG laws and for $c<\alpha-\frac12$ for the variance-gamma law with shape $\alpha$ \citep[Proposition~2 and Corollary~1]{delacruz:humeniy:2026}. Choosing $c$ small enough to satisfy all components at once and applying Corollary~\ref{cor:anydesign} proves the claim. \qed

\subsection{Proofs of Theorem~\ref{thm:block} and Corollary~\ref{cor:mvt}}\label{sec:proof-block}

We first record a bound for the fractional moments of a Gaussian vector.

\begin{lemma}\label{lem:chisq}
Let $\xi\sim\N_n(0,\Sigma)$ and $c\in(0,1]$. Then $\E\norm\xi^{2c}\le \vartheta_n(c)\,(\tr\Sigma)^c$, with $\vartheta_n(c)\le1$ as in Section~\ref{sec:shared}, and equality holds when $\Sigma$ is a multiple of $I_n$.
\end{lemma}

\begin{proof}
Let $\omega_1,\dots,\omega_n\ge0$ be the eigenvalues of $\Sigma$, so that $\norm\xi^2$ has the law of $\sum_i\omega_i\zeta_i^2$ with $\zeta_1,\dots,\zeta_n$ independent standard normal. The function $\phi(\omega)=\E(\sum_i\omega_i\zeta_i^2)^c$ is concave on $\R_+^n$, because $x\mapsto x^c$ is concave and the argument is linear in $\omega$, and it is symmetric, because the $\zeta_i$ are exchangeable. Averaging over the $n!$ permutations of $\omega$ and using concavity, $\phi(\omega)\le\phi(\bar\omega1_n)$ with $\bar\omega=\tr\Sigma/n$. Finally $\phi(\bar\omega1_n)=\bar\omega^c\E[(\chi^2_n)^c]=\bar\omega^c2^c\Gamma(c+\frac n2)/\Gamma(\frac n2)=\vartheta_n(c)(\tr\Sigma)^c$, and $\vartheta_n(c)\le1$ by Jensen's inequality.
\end{proof}

\emph{Proof of Theorem~\ref{thm:block}.} We indicate the changes to the proof of Theorem~\ref{thm:main}. Lemma~\ref{lem:tilted} holds for $H^{(n)}_s$ with $m_{n/2}$ in place of $m_{1/2}$: its proof in \citet[Lemma~5]{delacruz:humeniy:2026} uses only that $H_s$ has density proportional to $e^{-s\eta}$ with respect to a fixed finite measure, here $\eta^{n/2}H(d\eta)$. In particular $\E^{(n)}_s[\eta^c]\le m_{c+n/2}/m_{n/2}$. Lemma~\ref{lem:topology} holds with the same proof, with $\bar H_b=H_{b,d_b^\circ}$, $h_{b,d}=dH_{b,d}/d\bar H_b$ and $\varphi=\bigotimes_b\bar H_b\otimes\mathrm{Leb}_{p+q}$, the tilted factors being now $h_{b,d}(\eta_b)\eta_b^{n_b/2}e^{-s_b'\eta_b}/\mathcal Z^{(n_b)}_{H_{b,d}}(s_b')$, which, for a fixed destination, are dominated as before by a constant times $\exp(\sum_k\lambda_kb_k/2)$, since $\mathcal Z^{(n_b)}_{H_{b,d}}(s)\ge e^{-s(\eta_b+\delta)}\int_{(0,\eta_b+\delta)}t^{n_b/2}H_{b,d}(dt)>0$. Lemmas~\ref{lem:trace}, \ref{lem:means}, \ref{lem:blocklev} and~\ref{lem:traces} use only that $\Deta$ and $\Lambda_{\lambda,\eta}$ are positive diagonal matrices, which remains true with $\Deta=\diag(\eta_{b(1)},\dots,\eta_{b(q)})$, and the gamma full conditionals and \eqref{eq:lambdamoments} are unchanged. Instead of \eqref{eq:etamoments} we have, with $s_b'=\lambda_{k(b)}\norm{u_b'}^2/2$,
\begin{equation}\label{eq:etamoments-block}
\E[\eta_b^c\mid\lambda,\theta']\le M_c,\qquad\lambda_{k(b)}^{-c}\E[\eta_b^{-c}\mid\lambda,\theta']\le\kappa_b'2^{-c}\norm{u_b'}^{2c}+C_0\lambda_{k(b)}^{-c},
\end{equation}
where $\kappa_b'=\max_d\kappa^{(n_b)}_{H_{b,d}}(c)+\epsilon$, $M_c=\max m_{c+n_b/2}(H_{b,d})/m_{n_b/2}(H_{b,d})$ and $C_0$ are maxima over the pairs $(b,d)$ with $d\in\mathcal D_{k(b)}$, and $\epsilon>0$ is chosen so that
\[
\bar\gamma:=(1+\epsilon)^2\max_{b'}\kappa'_{b'}\frac{\Gamma(c+\frac{n_{b'}}2)}{\Gamma(\frac{n_{b'}}2)}\sigma_{b'}(c)<1,
\]
which is possible because $\bar\gamma\to\max_{b'}R_{b'}(c)\sigma_{b'}(c)<1$ as $\epsilon\to0$. We use
\[
v(\eta,\theta)=\alpha_1\sum_{b\in\mathcal B}\norm{u_b}^{2c}+\alpha_2\norm{\Deta^{1/2}u}^{2c}+\alpha_3\norm{y-W\theta}^{2c}+\sum_{b\in\mathcal B}(\eta_b^c+\eta_b^{-c})+\norm{\beta}^{2c},
\]
which contains the term $V_5=\norm\beta^{2c}$ of the proof of Theorem~\ref{thm:main}. Its sublevel sets are compact by the argument used for Theorem~\ref{thm:main}: on $\{v\le\bar v\}$ each $\eta_b$ lies in the compact set $\mathsf S_b\cap[\bar v^{-1/c},\bar v^{1/c}]$, $\norm{u_b}$ bounds $|u_i|$ for $i\in b$, and $\norm\beta^{2c}$ bounds $\beta$, whatever the rank of $X$. The terms $V_2$, $V_3$ and $V_5$ are handled as before. For $V_2$, the repetition of a weight within a block does not affect the bound, since $\E[\sum_i\eta_{b(i)}^c\mid\lambda,\theta']=\sum_bn_b\E[\eta_b^c\mid\lambda,\theta']\le qM_c$, which does not require independence of the weights. For $V_5$, Lemma~\ref{lem:means} bounds $\norm{\E(\theta\mid\lambda,\eta)}$ and $\Var(\beta\mid\lambda,\eta)\pe\Sigma_\beta$, so that $\E'V_5\le L_5$. For $V_4$, \eqref{eq:etamoments-block} and $\E'[\lambda_k^c]\le\zeta_k^c$ give $\E'V_4\le\bar\kappa2^{-c}\bar\zeta^cV_1'+L_4$.

For $V_1$, given $(\lambda,\eta)$ let $u_b\sim\N_{n_b}(m_b,\Sigma_b)$. By Lemma~\ref{lem:means}, $\norm{m_b}\le K_1$; writing $u_b=m_b+\xi_b$, the elementary inequality of Section~\ref{sec:proofs} and Lemma~\ref{lem:chisq} give $\E\norm{u_b}^{2c}\le(1+\epsilon)\vartheta_{n_b}(c)(\tr\Sigma_b)^c+C_{\epsilon,2c}K_1^{2c}$. With $\sigma_i^2=(Q_{\lambda,\eta}^{-1})_{ii}$, the decomposition used in the proof of Theorem~\ref{thm:main} gives $\sigma_i^2\le(1+\epsilon)\sum_j\Pi_{ij}^2(\lambda_{k(j)}\eta_{b(j)})^{-1}+C_{\epsilon,2}c_\lambda^{-1}\{(Z^TZ)^+\}_{ii}$. Summing over $i\in b$ and grouping $j$ by blocks,
\[
\tr\Sigma_b\le(1+\epsilon)\sum_{b'}F_{bb'}\,(\lambda_{k(b')}\eta_{b'})^{-1}+C_{\epsilon,2}\tau_Zc_\lambda^{-1},
\]
and, by $(\sum_jw_j)^c\le\sum_jw_j^c$ and $(1+\epsilon)^c\le1+\epsilon$,
\[
\E(V_1\mid\lambda,\eta)\le(1+\epsilon)^2\sum_{b'}\Big(\sum_b\vartheta_{n_b}(c)F_{bb'}^c\Big)\lambda_{k(b')}^{-c}\eta_{b'}^{-c}+C_8c_\lambda^{-c}+C_9 .
\]
By \eqref{eq:etamoments-block}, the coefficient of $\norm{u_{b'}'}^{2c}$ in $\E(V_1\mid\lambda,\theta')$ is $(1+\epsilon)^2\kappa_{b'}'2^{-c}\sum_b\vartheta_{n_b}(c)F_{bb'}^c$. Since $2^{-c}\vartheta_{n_{b'}}(c)n_{b'}^{c}=\Gamma(c+\frac{n_{b'}}2)/\Gamma(\frac{n_{b'}}2)$, this equals $(1+\epsilon)^2\kappa'_{b'}\Gamma(c+\frac{n_{b'}}2)\Gamma(\frac{n_{b'}}2)^{-1}\sigma_{b'}(c)\le\bar\gamma$. If $\delta_{k(b')}=0$, all the columns of $\Pi_Z$ that correspond to $b'$ vanish, so $F_{bb'}=0$ for every $b$ and block $b'$ does not contribute. By \eqref{eq:etamoments-block}, the remaining terms of $\E(V_1\mid\lambda,\theta')$ are therefore at most $C\sum_{k\in\mathcal K_+}\lambda_k^{-c}+C_8c_\lambda^{-c}+C_9$, and \eqref{eq:lambdamoments}, used only for $k\in\mathcal K_+$, together with $c_\lambda^{-c}\le\lambda_e^{-c}+\varsigma_{\max}^{2c}$, bounds their $\E'$-expectation by $A_{12}V_2'+A_{13}V_3'+L_1$. Hence $\E'V_1\le\bar\gamma V_1'+A_{12}V_2'+A_{13}V_3'+L_1$ with $\bar\gamma<1$, and the conclusion follows as in the proof of Theorem~\ref{thm:main}, including the case $\mathcal K_+=\emptyset$. \qed

\emph{Proof of Corollary~\ref{cor:mvt}.} Let $\Psi_g^{1/2}$ be the symmetric square root, and transform $v_b\mapsto\Psi_{g(b)}^{1/2}v_b$, $Z^{[b]}\mapsto Z^{[b]}\Psi_{g(b)}^{-1/2}$. As in the proof of Corollary~\ref{cor:scaling}, this bijection maps the model to model \eqref{eq:model-a}, \eqref{eq:model-block}, \eqref{eq:model-c} with blocks of size $n_0$ given by the subjects, components given by the groups and the same mixing laws, and it maps each full conditional to the corresponding one: the gamma rates depend on $\eta_bv_b^T\Psi_{g(b)}v_b=\eta_b\norm{\Psi_{g(b)}^{1/2}v_b}^2$, the tilting parameters on $\lambda_{g(b)}\norm{\Psi_{g(b)}^{1/2}v_b}^2/2$, and the update of $\theta$ is Gaussian and the map is linear. Hence the two Gibbs chains converge at the same rate. The transformation multiplies $Z$ on the right by an invertible matrix that is block diagonal with respect to the groups, so the ranks in \ref{C2}--\ref{C3} are unchanged. Because the rows of the subjects are disjoint, the null space of the transformed design is the direct sum over subjects of the null spaces of $Z^{[b]}\Psi^{-1/2}_{g(b)}$, so $\Pi_Z$ is block diagonal and $F_{bb'}=0$ for $b\ne b'$. The diagonal block $\Pi_b$ of $\Pi_Z$ is the orthogonal projection onto $\mathcal N(Z^{[b]}\Psi^{-1/2}_{g(b)})$, so $F_{bb}=\tr(\Pi_b^2)=\tr\Pi_b=\mathrm{nul}_b$. Thus $\sigma_b(c)=(\mathrm{nul}_b/n_0)^c$ and Theorem~\ref{thm:block} gives the first claim.

For $H_d=\Ga(d/2,d/2)$, the tilted law is $\Ga(A,\frac d2+s)$ with $A=\frac{d+n_0}2$, so $\E^{(n_0)}_s[\eta^{-c}]=\Gamma(A-c)\Gamma(A)^{-1}(\frac d2+s)^c$ for $c<A$, and $\kappa^{(n_0)}_{H_d}(c)=\Gamma(A-c)/\Gamma(A)$. For the slash law $H_\nu=\mathrm{Beta}(\nu,1)$ and $A=\nu+\frac{n_0}2$, the tilted law is $\mathrm{Beta}(A,1)$ for $s=0$ and the $\Ga(A,s)$ law truncated to $(0,1)$ for $s>0$, in which case $\E^{(n_0)}_s[\eta^{-c}]=s^c\Gamma(A-c)P(A-c,s)/\{\Gamma(A)P(A,s)\}$, so $\kappa^{(n_0)}_{H_\nu}(c)=\Gamma(A-c)/\Gamma(A)$ because $P(x,s)\to1$ as $s\to\infty$. In both cases all the moments $m_r$, $r>0$, are finite and $\log R^{(n_0)}_{H}(c)=\log\Gamma(c+\frac{n_0}2)-\log\Gamma(\frac{n_0}2)+\log\Gamma(A-c)-\log\Gamma(A)$ vanishes at $c=0$ with derivative $\psi(\frac{n_0}2)-\psi(A)<0$ there, since $A>\frac{n_0}2$. Hence $R^{(n_0)}_{H}(c)<1$ on an interval $(0,c_H)$ for each law $H$ of the grids, and, the grids being finite, some $c>0$ satisfies $R^{(n_0)}_{H}(c)(\mathrm{nul}_b/n_0)^c\le R^{(n_0)}_{H}(c)<1$ for all subjects and tail parameters at once. \qed

\section{Illustration: random intercepts and slopes for \texorpdfstring{$\beta$}{beta}-HCG data}\label{sec:app}

We use the $\beta$-HCG data analysed in \citet{delacruz:humeniy:2026}: $N=375$ measurements of $\log_{10}\beta$-HCG on 173 women attending a fertility clinic \citep{MarshallBaron2000}, 124 with a normal pregnancy and 49 with an abnormal outcome. The fixed effects are those of \citet{delacruz:humeniy:2026}: an intercept, time $t$ in weeks, $t^2$, the group indicator and its interaction with time, so $p=5$; $\beta_4$ is the difference between the groups at $t=0$. That paper found that the main departure from a normal random-intercept model with a common scale is a larger between-subject variability among abnormal pregnancies. Here we add a random slope in time for each woman and let the four components (intercepts and slopes, of normal and abnormal pregnancies) have their own scales and their own unknown degrees of freedom. Thus $Z=(Z_1\ Z_1\circ t)$, with $Z_1$ the subject indicator matrix and $q=346$, split into $K=4$ blocks.

\paragraph{The conditions.} Since 52 women have a single measurement, $r=\rank(Z)=294$. Table~\ref{tab:cond} gives the quantities in Theorem~\ref{thm:main}. The conditions \ref{C2}--\ref{C3} on the gamma shapes hold for any positive values. The intercepts of women with a single observation at time $t$ have $\ell_i=t^2/(1+t^2)$, and since these times range from $2$ to $11.4$ weeks, $\ell_k^*$ is $0.99$ for the intercept components. With $c=1$, Proposition~\ref{prop:families}(a) therefore requires $\min\mathcal D_k>1.99$ for the intercepts. Optimizing over $c$ gives the smaller thresholds in the column ``best $c$'' of Table~\ref{tab:cond}, and with a common $c$ for all components the requirement is $\min\mathcal D_k>1.32$. The null space of $Z$ consists of 52 disjoint pairs, so Example~\ref{ex:pairs} applies, and the symmetric rescaling with $c=\frac12$ gives $\min\mathcal D_k>1$. Finally, the 52 intercepts of the women with a single observation form a pivot set, and Corollary~\ref{cor:families} shows that the chain is geometrically ergodic for every finite grid. For comparison, bounding the rank-deficient part of the trace by the rank deficiency $q-r$, as the trace inequality of \citet[Lemma~3]{roma:hobe:2014} does, leads with $c=1$ to $d>1+q-r=53$.

\begin{table}[t]
\centering\small
\caption{$\beta$-HCG data with random intercepts and slopes: quantities entering Theorem~\ref{thm:main} for each component, and the resulting lower bound on the degrees of freedom of Student-$t$ random effects: with $c=1$ ($d>1+\ell_k^*$), with the best $c\in(0,1]$ for each component, and with the symmetric rescaling of Example~\ref{ex:pairs} and $c=\frac12$. By Corollary~\ref{cor:families}, every $d>0$ is in fact allowed. $L_k=\sum_{i:k(i)=k}\ell_i$.}\label{tab:cond}
\begin{tabular}{lrrrrrrrr}
\toprule
Component & $q_k$ & $r_{-k}$ & $\delta_k$ & $L_k$ & $\ell_k^*$ & $c=1$ & best $c$ & symmetric\\
\midrule
Intercepts, normal & 124 & 205 & 35 & 33.5 & 0.992 & 1.992 & 1.314 & 1\\
Intercepts, abnormal & 49 & 262 & 17 & 15.8 & 0.992 & 1.992 & 1.317 & 1\\
Slopes, normal & 124 & 205 & 35 & 1.5 & 0.109 & 1.109 & 0.197 & 1\\
Slopes, abnormal & 49 & 262 & 17 & 1.2 & 0.200 & 1.200 & 0.445 & 1\\
\bottomrule
\end{tabular}
\end{table}

\paragraph{Models and priors.} We fitted four models: M0, normal random intercepts with group-specific scales, which is the model preferred in \citet{delacruz:humeniy:2026}; M1, normal random intercepts and slopes with four scales; M2, the same with Student-$t$ random effects and unknown degrees of freedom $d_k$ for each component; and M3, Student-$t$ random intercepts with two group-specific scales and unknown $d_k$. The priors are $\beta\sim\N_5(0,1000I)$, $\lambda_e\sim\Ga(1,0.1)$, $\lambda_k\sim\Ga(1,0.1)$ for the intercept components, and $\lambda_k\sim\Ga(1,0.1/\overline{t^2})$ for the slope components, where $\overline{t^2}=27.4$ is the mean of $t^2$ over the measurements. The last choice gives the precision of $(\overline{t^2})^{1/2}$ times a slope the same prior as the precision of an intercept, and makes the prior on the random-effect scales invariant to the unit of time. The $d_k$ have independent uniform priors on the grid $\{2,2.5,3,4,5,6,8,10,15,20,30,50,100\}$, under which $\Pr(d_k\le4)=0.31$ and $\Pr(d_k\le10)=0.62$. Theorem~\ref{thm:main} and Corollary~\ref{cor:families} give geometric ergodicity for the four models; for M2 and M3 this also follows from Proposition~\ref{prop:families} with $c=1$, since $\min\mathcal D_k=2>1+\ell_k^*$.

\paragraph{Output analysis.} For each model we ran four chains of $30{,}000$ iterations from dispersed starting values and discarded the first $5{,}000$ of each. We monitored $g=(\beta^T,\log\sigma_e,\log\sigma_1,\dots,\log\sigma_K)^T$, with $\sigma_e=\lambda_e^{-1/2}$ and $\sigma_k=\lambda_k^{-1/2}$. By Proposition~\ref{prop:moments}, $g$ has finite posterior moments of all orders, and geometric ergodicity yields a CLT for $g$ and strong consistency of the multivariate batch-means estimator of its asymptotic covariance with batch size $\lfloor n^{0.6}\rfloor$. The multivariate effective sample size per iteration was between $0.29$ and $0.37$ for M1 and M2 and between $0.50$ and $0.78$ for M0 and M3, so the pooled sample of each model exceeds by a wide margin $\mathrm{minESS}(10,0.05,0.05)=8831$ for M1 and M2, where $g$ has ten components, and $\mathrm{minESS}(8,0.05,0.05)=8804$ for M0 and M3. The largest Monte Carlo standard error of a component of $g$ was $0.013$, and the split potential scale reduction factors over the eight half-chains were at most $1.0009$. A run of M2 took about two and a half minutes.

\paragraph{Model comparison.} The likelihood of a woman's measurements, with her random effects and weights integrated out, is a mixture of normal densities over the weights; we computed it by Gauss--Laguerre quadrature with $40$ nodes per weight, which agrees with adaptive integration to a relative error below $2\times10^{-6}$. From it we obtained WAIC and LPML with the subject as the unit, and standard errors of WAIC and of WAIC differences from the subject-level contributions \citep{vehtari:gelman:gabry:2017}.

\begin{table}[t]
\centering\small
\caption{$\beta$-HCG data: model comparison and posterior summaries. M0: normal random intercepts, two scales; M1: normal random intercepts and slopes, four scales; M2: Student-$t$ random intercepts and slopes with unknown $d_k$, four scales; M3: Student-$t$ random intercepts with unknown $d_k$, two scales. Standard errors in parentheses; $\Delta$WAIC is the difference with M0. $\beta_4$ is the group difference at $t=0$, and $\beta_4+4\beta_5$ at $t=4$ weeks; posterior means with 95\% intervals.}\label{tab:fit}
\begin{tabular}{lrrrcc}
\toprule
Model & WAIC & $\Delta$WAIC & LPML & $\beta_4$ & $\beta_4+4\beta_5$\\
\midrule
M0 & 465.4 (41.4) & 0 & $-232.7$ & $0.20\ (-0.15,0.53)$ & $-0.49\ (-0.70,-0.29)$\\
M1 & 471.5 (41.4) & 6.1 (4.6) & $-236.0$ & $0.24\ (-0.10,0.58)$ & $-0.50\ (-0.71,-0.29)$\\
M2 & 472.9 (41.5) & 7.5 (5.1) & $-236.5$ & $0.27\ (-0.08,0.60)$ & $-0.47\ (-0.68,-0.25)$\\
M3 & 466.8 (41.6) & 1.4 (1.6) & $-233.4$ & $0.23\ (-0.12,0.57)$ & $-0.46\ (-0.67,-0.25)$\\
\bottomrule
\end{tabular}
\end{table}

\begin{table}[t]
\centering\small
\caption{$\beta$-HCG data: posterior of the degrees of freedom $d_k$. Under the uniform prior on the grid, the median is $8$, $\Pr(d_k\le4)=0.31$ and $\Pr(d_k\le10)=0.62$. The Monte Carlo standard errors of the posterior probabilities, computed by batch means from every tenth draw of the four chains, are at most $0.005$.}\label{tab:dof}
\begin{tabular}{llrrr}
\toprule
Model & Component & Posterior median & $\Pr(d_k\le4\mid y)$ & $\Pr(d_k\le10\mid y)$\\
\midrule
M2 & Intercepts, normal & 20 & 0.04 & 0.31\\
M2 & Intercepts, abnormal & 10 & 0.21 & 0.54\\
M2 & Slopes, normal & 20 & 0.03 & 0.27\\
M2 & Slopes, abnormal & 8 & 0.28 & 0.61\\
M3 & Intercepts, normal & 20 & 0.04 & 0.31\\
M3 & Intercepts, abnormal & 10 & 0.20 & 0.53\\
\bottomrule
\end{tabular}
\end{table}

\paragraph{Results.} Tables~\ref{tab:fit} and~\ref{tab:dof} summarize the fits. WAIC and LPML give the same ranking. Neither criterion gives evidence in favour of random slopes: M1 is worse than M0 by $6.1$ units of WAIC, with a standard error of $4.6$, and under M1 the posterior means of the scales $\sigma_k$ of the slopes are small ($0.03$ and $0.08$ per week for normal and abnormal pregnancies, against $0.17$ and $0.54$ for the intercepts). Nor do they give evidence in favour of heavy tails: replacing normal by Student-$t$ random effects increases WAIC by $1.4$ (standard error $1.6$) for the intercept models and by $1.4$ ($1.1$) for the intercept and slope models. Accordingly, the posteriors of the $d_k$ move towards large values for the components with many women and remain close to the prior for the intercepts of abnormal pregnancies and for their slopes, which are informed by 49 women, many of them with one or two measurements. These results agree with the finding of \citet{delacruz:humeniy:2026} that the main departure from a common-scale normal model is the larger between-subject variability of abnormal pregnancies. The conclusions on the groups do not depend on the model: the difference at $t=0$ is uncertain in all four, with $\Pr(\beta_4>0\mid y)$ between $0.87$ and $0.94$, and at four weeks the abnormal group is lower by about $0.5$ on the $\log_{10}$ scale, with intervals that exclude zero. The posterior of each $d_k$ is a by-product of the sampler, and its Monte Carlo error is controlled by the same theory, since the indicator functions $\mathbf 1\{d_k\le d_0\}$ are bounded and the CLT applies to them under geometric ergodicity \citep{Jones2004}.

\paragraph{Sensitivity to the priors on the scales.} With so many women observed once or twice, the scales of the slopes are weakly identified and their prior matters. With $\Ga(1,0.1)$ for all the precisions, whose density decays fast for large precisions, the posterior means of the scales of the slopes are pushed up to $0.07$ and $0.15$, and the values of WAIC for M0, M1, M2 and M3 are $465.4$, $514.5$, $516.7$ and $466.8$; M1 is worse than M0 by $49.1$ units (standard error $12.3$). With $\Ga(1,0.001)$ for all the precisions, the posterior mean of the scale of the intercepts of normal pregnancies drops from $0.17$ to $0.07$, the values of WAIC are $465.6$, $472.8$, $473.9$ and $467.4$, and the mixing of the chains deteriorates, with multivariate effective sample sizes per iteration of about $0.15$ and split potential scale reduction factors up to $1.05$ for the intercept and slope models; the figures for this prior are therefore less precise than the others. The ranking of the models and the group difference at four weeks are the same under the three priors, the posterior mean of $\beta_4+4\beta_5$ lying between $-0.50$ and $-0.45$ in all fits. The difference at $t=0$ is more sensitive, as expected for an extrapolation: $\Pr(\beta_4>0\mid y)$ ranges from $0.86$ to $0.98$ over the models and priors. Geometric ergodicity holds under all three priors, but it provides no quantitative bound on the efficiency of the sampler, which here deteriorates for very diffuse priors.

\section{Shared weights: exploration of the scale and the tail parameter}\label{sec:sim}

How precisely can the scale of the random effects and the tail parameter be estimated by Monte Carlo when the weights are shared and the tails are very heavy? To examine this question in a setting covered by Corollary~\ref{cor:mvt}, we simulated $m=60$ subjects, $15$ of them with a single observation and the others with two to six, from $y_{ij}=\beta_1+\beta_2t_{ij}+v_{i1}+v_{i2}t_{ij}+e_{ij}$, with $t_{ij}$ uniform on $(0,1)$, $\beta=(1,0.5)^T$, $\sigma_e=0.3$ and $v_i\mid\eta_i\sim\N_2(0,0.25\,\eta_i^{-1}I)$, $\eta_i\sim\Ga(\frac12,\frac12)$, so that the random-effect vectors are bivariate Cauchy. Thus $N=201$, $q=120$ and $\rank(Z)=105$, each subject with a single observation carrying a one-dimensional part of the null space ($\mathrm{nul}_b/n_0=\frac12$). We fitted model \eqref{eq:mvt} with one group, $\Psi=I$, a uniform prior for $d$ on $\{0.5,1,2,4,8,16\}$, $\beta\sim\N_2(0,1000I)$ and $\Ga(1,0.1)$ priors for both precisions. By Corollary~\ref{cor:mvt} the Gibbs sampler is geometrically ergodic. We ran four chains of $105{,}000$ iterations from dispersed starting values and discarded the first $5000$ of each.

The posterior of $d$ concentrates on the two heaviest values of the grid, with $\Pr(d=0.5\mid y)=0.355$ and $\Pr(d=1\mid y)=0.645$. The split $\hat R$ statistics for $g=(\beta_1,\beta_2,\log\sigma_e,\log\sigma_u)^T$ are at most $1.0006$. The effective sample size per iteration is between $0.26$ and $0.41$ for $\beta_1$, $\beta_2$ and $\log\sigma_e$, but only $0.021$ for $\log\sigma_u$ and $0.030$ for the indicator $\mathbf 1\{d=0.5\}$. The multivariate effective sample size of $g$ per iteration is about $0.2$, so that, extrapolating as $8431/(\mathrm{mESS}/n)$, a single chain needs about $43{,}000$ iterations to reach $\mathrm{minESS}(4,0.05,0.05)=8431$, and the Monte Carlo standard error of $\Pr(d=0.5\mid y)$ from the four chains is $0.004$.

The contrast between the multivariate effective sample size and that of $\log\sigma_u$ and of the tail indicator motivates a small experiment on how the exploration of these quantities depends on the information per subject. With the same model and grid, we varied the number $n_1\in\{0,15,30\}$ of the $60$ subjects with a single observation, so that $\rank(Z)=120-n_1$, and the true degrees of freedom $d_0\in\{1,4\}$. For each of the six configurations we generated three data sets and ran two chains of $55{,}000$ iterations, discarding the first $5000$. Table~\ref{tab:info} reports the results. Corollary~\ref{cor:mvt} applies to all of them.

\begin{table}[t]
\centering\footnotesize\setlength{\tabcolsep}{2.5pt}
\caption{Shared-weight random intercepts and slopes, $60$ subjects of which $n_1$ have a single observation, true degrees of freedom $d_0$, uniform prior for $d$ on $\{0.5,1,2,4,8,16\}$; three data sets per row, two chains of $50{,}000$ iterations after burn-in per data set. ESS per iteration: range over data sets and over $\beta_1,\beta_2$ for the regression coefficients, averages over data sets for $\log\sigma_e$ and $\log\sigma_u$; mESS$/n$ for $g=(\beta_1,\beta_2,\log\sigma_e,\log\sigma_u)^T$; range over data sets of $\Pr(d\le1\mid y)$, of the ESS per iteration of $\mathbf 1\{d\le1\}$ for the data sets in which this probability lies in $(0.02,0.98)$, and the largest Monte Carlo standard error of the probability and the largest split $\hat R$ of $g$.}
\label{tab:info}
\begin{tabular}{rrrcccccccc}
\toprule
 & & & \multicolumn{3}{c}{ESS per iteration} & & & & & \\
\cmidrule(lr){4-6}
$d_0$ & $n_1$ & $\rank(Z)$ & $\beta$ & $\log\sigma_e$ & $\log\sigma_u$ & mESS$/n$ & $\Pr(d\le1\mid y)$ & ESS $\mathbf 1\{d\le1\}$ & MCSE & $\max\hat R$\\
\midrule
1 & 0 & 120 & 0.33--0.55 & 0.35 & 0.057 & 0.26 & 0.18--0.99 & 0.06--0.13 & 0.006 & 1.0004\\
1 & 15 & 105 & 0.29--0.65 & 0.34 & 0.042 & 0.24 & 0.95--1.00 & 0.18 & 0.002 & 1.0009\\
1 & 30 & 90 & 0.34--0.64 & 0.29 & 0.049 & 0.23 & 0.92--1.00 & 0.08 & 0.003 & 1.0013\\
\addlinespace
4 & 0 & 120 & 0.47--0.82 & 0.36 & 0.092 & 0.36 & 0.00--0.01 & -- & 0.001 & 1.0004\\
4 & 15 & 105 & 0.46--0.74 & 0.30 & 0.084 & 0.33 & 0.00 & -- & 0.001 & 1.0003\\
4 & 30 & 90 & 0.47--0.79 & 0.28 & 0.068 & 0.31 & 0.00--0.01 & -- & 0.001 & 1.0001\\
\bottomrule
\end{tabular}
\end{table}

Three features stand out. First, in every configuration the scale of the random effects mixes far more slowly than the regression coefficients: its effective sample size per iteration is $0.04$ to $0.09$, against $0.29$ to $0.82$ for $\beta$, and it is lower for the heavier tails. The multivariate effective sample size of $g$, between $0.23$ and $0.36$ per iteration, does not reveal this: it summarizes joint precision through the volume of a confidence region, and the multivariate criterion does not ensure a prescribed Monte Carlo precision for each component; nor does it involve the tail probabilities, whose indicators are not part of $g$. Second, in these realizations the contrast between quantities is larger than the differences between configurations: for $d_0=4$ the effective sample size of $\log\sigma_u$ falls from $0.092$ to $0.068$ as $n_1$ goes from $0$ to $30$, but for $d_0=1$ it does not change monotonically. With three data sets per configuration, in which the information, the rank and the data change together, the experiment does not support a general statement on the effect of the number of observations per subject. Third, the precision with which the tail is explored depends on how uncertain the posterior of $d$ is. For intermediate values of $\Pr(d\le1\mid y)$ the effective sample size per iteration of the indicator can be as low as $0.06$; even then, $100{,}000$ draws give an estimated Monte Carlo standard error below $0.006$ for this probability. When the probability is close to $0$ or $1$ its estimated standard error is small, but this concerns only the event $\{d\le1\}$ and does not show that the other values of the grid are well explored.

For Bayesian practice the theory and these results lead to a simple recommendation. Since the chain is geometrically ergodic and the quantities involved are either bounded or have posterior moments of all orders, Monte Carlo standard errors are valid for the regression coefficients, the logarithms of the scales and the posterior probabilities of the tail parameters, and they should be computed and reported for each of these quantities separately. A run that is long enough for the regression coefficients, or for a multivariate criterion that sets no target for each component, may be several times too short for the scale of the random effects and for statements about the tails, which are often the quantities of interest in a robust analysis.

\section{Discussion}\label{sec:disc}

We have extended the analysis of the block Gibbs sampler for linear mixed models with scale mixtures of normal random effects to several components, rank-deficient designs, unknown tail parameters on finite grids and shared weights. The new ingredients are the component-wise control of the rank deficiency through $\delta_k$ and Lemma~\ref{lem:blocklev}, and the block argument of Theorem~\ref{thm:block}; the rescaling of the design, which completes the results for the usual families, is the tool of \citet{delacruz:humeniy:2026} adapted to this setting. The conditions on the mixing distributions are expressed through the projection onto the null space of the design, which can be computed before running the sampler, and Corollaries~\ref{cor:scaling} and~\ref{cor:anydesign} show that they may be checked after rescaling the columns of the design. With individual weights, a suitable rescaling makes them hold for all the usual families and any design, so that the rank of the design restricts only the shapes of the gamma priors on the precision parameters, through \ref{C2} and \ref{C3}. With shared weights, Theorem~\ref{thm:block} keeps a geometric condition on the blocks of the null space, and the result for any finite grid of Student-$t$ and slash tail parameters is obtained for the subject-specific designs of Corollary~\ref{cor:mvt}.

Three extensions seem natural. First, Corollary~\ref{cor:mvt} covers multivariate scale mixtures with a known shape matrix and an unknown scale. With an unknown, unstructured precision matrix $\Xi$ and a Wishart prior with scale matrix $S_0$, the sampler keeps its three-block structure, but the drift argument does not carry over. In a direction of the null space of $Z$, the conditional variance of a subject's random effect depends on $\Xi$ in that direction, while the shared weight is tilted by the quadratic form in $\Xi$ along the current random effect; the two no longer cancel as the scales do in the proof of Theorem~\ref{thm:main}, and one-step bounds produce products of terms of the drift function. For multivariate regression with errors that are scale mixtures of normals, which share the weight structure but have no unidentified directions, \citet{RoyHobert2010} and \citet{HobertJungKhareQin2018} measure the residuals in the metric of the current scale matrix, which then cancels. An analogous metric here would depend on the state of the chain, through the shared weights and the current random effects, and the difficulty is to control the change of this metric from one sweep to the next. Second, a continuous prior on the tail parameters breaks the conditional conjugacy of the joint update of the tail parameters and the weights and may require an additional nonconjugate update, such as a Metropolis--Hastings step; \citet{JonesRobertsRosenthal2014} give conditions under which conditional Metropolis--Hastings samplers inherit geometric ergodicity from a corresponding exact Gibbs sampler. Our sampler draws $\lambda$ and then $(d,\eta)$ from their full conditionals, which is not in general an exact draw from the joint conditional law of $(\lambda,d,\eta)$ given $\theta$, so Theorem~\ref{thm:main} does not by itself provide the property required there; establishing it, or an analogous result for the three-block structure, is an open problem. Third, heavy-tailed errors, with latent weights for the observations, are common in practice and are supported by our implementation, but they introduce a second set of weights in the conditional variance of $\theta$, which the present argument does not control. Our conditions are sufficient and not necessary. In particular, Lemma~\ref{lem:blocklev} is used with $\Lambda$ depending on the weights, and a bound that exploits the specific structure of $\Deta^{1/2}\mathcal N(Z)$ might weaken \ref{C3}, although this condition is not binding in our application.

\section*{Acknowledgements}
{\sloppy The work of the first author was partially supported by ANID, Chile, through grants ANID/\allowbreak FONDECYT/\allowbreak 1260228, ANID/\allowbreak ANILLO/\allowbreak ACT210096 and ANID-MILENIO-NCN2024\_103, and by the Data Observatory Foundation, ANID Technology Center No.~DO210001. The work of the second author was partially supported by an ANID PhD scholarship.\par}

\section*{Code and data availability}
Python code implementing the Gibbs sampler with several components, tail parameters on finite grids and shared weights, the verification tests of Appendix~\ref{app:code}, the analyses of Section~\ref{sec:app} and the simulations of Section~\ref{sec:sim} is available at \url{https://github.com/rolandodelacruz74/smnlmm-components}. The $\beta$-HCG data are available from the first author upon request.

\appendix
\section{Verification of the implementation}\label{app:code}

The sampler is implemented in Python, with any number of components, component-specific mixing families, tail parameters on finite grids drawn jointly with the weights, and an exact draw of $\theta$ (through the Schur complement when every row of $Z$ has at most one non-zero entry, and through a Cholesky factorization of $P_{\lambda,\eta}$ otherwise). The tilted laws and the normalizing functions $\mathcal Z_{H_d}$ were compared with adaptive quadrature. The joint draw of the tail parameter and the weights was checked against the exact conditional law on the grid. We also ran the joint distribution test of \citet{Geweke2004} with $N=9$ observations on five configurations: random intercepts and slopes for four subjects, one of them with a single observation (so that $Z$ is rank deficient), with Student-$t$ random intercepts with $d$ unknown on $\{2.5,4,8\}$ and Student-$t$ random slopes with $d=3$; slash random intercepts with $\nu$ unknown on $\{0.8,1.5,3\}$ and on $\{1.5,3,6\}$, combined with Student-$t$ errors with unknown degrees of freedom; Student-$t$ random intercepts with contaminated normal errors; and slash random intercepts with $\nu=0.8$. The configurations with heavy-tailed errors check parts of the implementation outside the setting studied here. The priors were $\beta\sim\N_2(0,I)$ and $\Ga(3,2)$ for all the precisions. In each configuration we compared, over $60{,}000$ iterations, the means of $\beta_1$, $\beta_2$, $\lambda_e$, the first and last component precisions, $|u_1|^{1/2}$, $|u_q|^{1/2}$, the logarithms of a latent weight of each kind and, where applicable, the tail parameters, 48 comparisons in all. The random effects themselves were not compared, because for the slash law with $\nu=0.8$ they have infinite prior variance; $|u_i|^{1/2}$ has finite prior variance in all configurations, since $\E[\eta^{-1/2}]<\infty$ for every mixing law used and $\E[\lambda_k^{-1/2}]<\infty$. Standardized by batch-means standard errors, which serve as a practical scale for this empirical check, the largest of the 48 differences was $2.99$ in absolute value. The code is available at \url{https://github.com/rolandodelacruz74/smnlmm-components}.

\section{Proofs of Lemmas~\ref{lem:trace}--\ref{lem:means} and of the elementary facts}\label{app:lemmas}

These proofs are those of \citet{delacruz:humeniy:2026}, written for the notation of this paper.

\emph{Lemma~\ref{lem:trace}.} (a) Since $Q\succeq\Omega\succeq\Upsilon$, $Q^{-1}\pe\Omega^{-1}\pe\Upsilon^{-1}$, which gives the second bound. For the first, it suffices to prove $P_A\Omega^{-1}P_A\pe\upsilon^{-1}(A^TA)^+$, because $Q^{-1}\pe\Omega^{-1}$. Let $k=\rank(A)$. If $k=0$, then $P_A=0$; if $k=n$, then $P_A=I$ and $\Omega\succeq\upsilon A^TA$, which is invertible. Otherwise let $O=(O_1\ O_2)$ be an orthogonal matrix whose first $k$ columns span the row space of $A$ and whose last $n-k$ columns span $\mathcal N(A)$, so that $P_A=O_1O_1^T$ and $(A^TA)^+=O_1B_A^{-1}O_1^T$ with $B_A=O_1^TA^TAO_1$ positive definite. With $\Upsilon_{ab}=O_a^T\Upsilon O_b$,
\[
O^T\Omega O=\begin{pmatrix}\upsilon B_A+\Upsilon_{11}&\Upsilon_{12}\\ \Upsilon_{21}&\Upsilon_{22}\end{pmatrix},
\]
and the formula for the inverse of a partitioned matrix gives $O_1^T\Omega^{-1}O_1=(\upsilon B_A+\Upsilon_{11}-\Upsilon_{12}\Upsilon_{22}^{-1}\Upsilon_{21})^{-1}$. The Schur complement $\Upsilon_{11}-\Upsilon_{12}\Upsilon_{22}^{-1}\Upsilon_{21}$ of the positive definite matrix $O^T\Upsilon O$ is positive definite, so $O_1^T\Omega^{-1}O_1\pe(\upsilon B_A)^{-1}$ and $P_A\Omega^{-1}P_A=O_1(O_1^T\Omega^{-1}O_1)O_1^T\pe\upsilon^{-1}(A^TA)^+$.
(b) Since $P_A+\Pi_A=I$ and $\tr(P_A\Omega^{-1}\Pi_A)=\tr(\Omega^{-1}\Pi_AP_A)=0$, $\tr(\Omega^{-1})=\tr(P_A\Omega^{-1}P_A)+\tr(\Pi_A\Omega^{-1}\Pi_A)$, and part (a) bounds the two terms by $\upsilon^{-1}\tr\{(A^TA)^+\}$ and $\tr(\Pi_A\Upsilon^{-1}\Pi_A)=\sum_i\tilde\ell_i\upsilon_i^{-1}$, where the last equality uses $\Pi_A^2=\Pi_A$.
(c) Let $\upsilon_{\min}=\min_i\upsilon_i$ and let $\vartheta_1,\dots,\vartheta_n$ be the eigenvalues of $A^TA$. Since $\Omega\succeq\upsilon A^TA+\upsilon_{\min}I$,
\[
\tr(A\Omega^{-1}A^T)\le\tr\{A(\upsilon A^TA+\upsilon_{\min}I)^{-1}A^T\}=\sum_{i:\vartheta_i>0}\frac{\vartheta_i}{\upsilon\vartheta_i+\upsilon_{\min}}\le\frac{\rank(A)}{\upsilon}.
\]

\emph{Lemma~\ref{lem:tilted}.} If $0\le s<s'$, the density of $H_{s'}$ with respect to $H_s$ is $w/\E_s[w]$ with $w(\eta)=e^{-(s'-s)\eta}$, which is non-increasing. For a bounded non-decreasing $g$,
\[
\E_{s'}[g]-\E_s[g]=\frac{\E_s[gw]-\E_s[g]\E_s[w]}{\E_s[w]}\le0,
\]
because a non-decreasing and a non-increasing function of the same random variable have non-positive covariance (Chebyshev's association inequality). By monotone convergence, applied to $\min(g,n)$, the same holds for every non-decreasing $g\ge0$, and the reverse inequality holds for every non-increasing $g\ge0$. (a) Taking $g(\eta)=\eta^c$, $\sup_{s\ge0}\E_s[\eta^c]=\E_0[\eta^c]=m_{c+1/2}(H)/m_{1/2}(H)$. (b) Choose $S>0$ with $\E_s[\eta^{-c}]\le\kappa's^c$ for $s\ge S$, which is possible by the definition of $\kappa_H(c)$ as a $\limsup$. Since $\eta^{-c}$ is non-increasing, $\E_s[\eta^{-c}]\le\E_S[\eta^{-c}]\le\kappa'S^c=:C$ for $0\le s<S$.

\emph{Lemma~\ref{lem:means}.} Since $\theta\mid\lambda,\eta$ is normal, its mean minimizes
\[
\lambda_e\norm{y-W\theta}^2+u^T\Lambda_{\lambda,\eta}u+(\beta-\mu_\beta)^T\Sigma_\beta^{-1}(\beta-\mu_\beta)=\norm{D^{1/2}(\tilde y-\tilde A\theta)}^2,
\]
where
\[
\tilde A=\begin{pmatrix}X&Z\\0&I_q\\\Sigma_\beta^{-1/2}&0\end{pmatrix},\qquad\tilde y=\begin{pmatrix}y\\0\\\Sigma_\beta^{-1/2}\mu_\beta\end{pmatrix},\qquad D=\diag(\lambda_e1_N,\Lambda_{\lambda,\eta},1_p).
\]
The last $q+p$ rows of $\tilde A$ form an invertible matrix, so $\tilde A$ has full column rank and does not depend on $(\lambda,\eta)$, while $D$ is a positive diagonal matrix. Hence $\E(\theta\mid\lambda,\eta)=(\tilde A^TD\tilde A)^{-1}\tilde A^TD\tilde y=\tilde A^+\{\tilde A(\tilde A^TD\tilde A)^{-1}\tilde A^TD\}\tilde y$, with $\tilde A^+=(\tilde A^T\tilde A)^{-1}\tilde A^T$. By the theorem of \citet{Stewart1989} and \citet{OLeary1990}, $\chi=\sup_D\norm{\tilde A(\tilde A^TD\tilde A)^{-1}\tilde A^TD}_2<\infty$, the supremum being over all positive diagonal matrices $D$. Therefore $\norm{\E(\theta\mid\lambda,\eta)}\le K_1:=\norm{\tilde A^+}_2\chi\norm{\tilde y}$ and $\norm{y-W\E(\theta\mid\lambda,\eta)}\le K_2:=\norm y+\norm W_2K_1$. The argument uses only that $\Lambda_{\lambda,\eta}$ is a positive diagonal matrix, so the same constants apply in the models with shared weights.

\emph{Elementary facts.} For $p\le1$, $t\mapsto t^p$ is subadditive and one can take $C_{\epsilon,p}=1$; for $p>1$ and $t\in(0,1)$, convexity gives $(x+y)^p=\{(1-t)\frac{x}{1-t}+t\frac yt\}^p\le(1-t)^{1-p}x^p+t^{1-p}y^p$, and $t$ is chosen with $(1-t)^{1-p}\le1+\epsilon$. If $Y\sim\N(m,\sigma^2)$, write $Y=m+\sigma\xi$ with $\xi\sim\N(0,1)$; the first fact with $p=2c$ gives $|Y|^{2c}\le(1+\epsilon)\sigma^{2c}|\xi|^{2c}+C_{\epsilon,2c}|m|^{2c}$, and taking expectations gives the bound. The inequality $(\sum_jw_j)^c\le\sum_jw_j^c$ is the subadditivity of $t\mapsto t^c$, and Jensen's inequality applies because $t\mapsto t^c$ is concave.

\bibliographystyle{plainnat}
\bibliography{paper2}
\end{document}